\documentclass[twocolumn]{aastex631}
\usepackage{amsmath}

\begin{document}

\title{ALMA Observations of the Extraordinary Carina Pillars: A Complementary Sample}

\author{Geovanni Cortes-Rangel}
\affiliation{Instituto de Radioastronom\'\i a y Astrof\'\i sica, Universidad Nacional Aut\'onoma de M\'exico, \\P.O. Box 3-72, 58090, Morelia, Michoac\'an, M\'exico.}

\author{ Luis A. Zapata}
\affiliation{Instituto de Radioastronom\'\i a y Astrof\'\i sica, Universidad Nacional Aut\'onoma de M\'exico, \\P.O. Box 3-72, 58090, Morelia, Michoac\'an, M\'exico.}

\author{Pedro R. Rivera-Ortiz}
\affiliation{Instituto de Radioastronom\'\i a y Astrof\'\i sica, Universidad Nacional Aut\'onoma de M\'exico, \\P.O. Box 3-72, 58090, Morelia, Michoac\'an, M\'exico.}

\author{Megan Reiter}
\affiliation{Department of Physics and Astronomy, Rice University, 6100 Main St - MS 108, Houston, TX 77005, USA}

\author{Satoko Takahashi}
\affiliation{National Astronomical Observatory of Japan, 2-21-1 Osawa, Mitaka, Tokyo 181-8588, Japan}
\affiliation{Department of Astronomical Science, The Graduate University for Advanced Studies (SOKENDAI), 2-21-1, Osawa, Mitaka, Tokyo 181-8588, Japan}

\author{Josep M. Masqué}
\affiliation{Departamento de Astronomía, Universidad de Guanajuato, Apartado Postal 144, 36000, Guanajuato, Guanajuato, México}




\begin{abstract}




We present a study of six dusty and gaseous pillars (containing the HH 1004 and HH 1010 objects) and globules (that contain the HH 666, HH 900, HH 1006, and HH 1066 objects) localized in the Carina nebula using sensitive and high angular resolution ($\sim$0.3$''$) Atacama Large Millimeter/Sub-millimeter Array (ALMA) observations. This is a more extensive study that the one presented in \citet{Cortes}. As in this former study, we also analyzed the 1.3 mm continuum emission and C$^{18}$O(2$-$1), N$_2$D$^+$(3$-$2) and $^{12}$CO(2$-$1) spectral lines. These new observations revealed the molecular outflows emanating from the pillars, the dusty envelopes$+$disks that are exciting them, and the extended HH objects far from their respective pillars. We reveal that the masses of the disks$+$envelopes are in a range of 0.02 to 0.38 M$_\odot$, and those for the molecular outflows are of the order of 10$^{-3}$ M$_\odot$, which suggests that their exciting sources might be low- or intermediate-mass protostars as already revealed in recent studies at infrared and submillimeter bands. In the regions associated with the objects HH 900 and HH 1004, we report multiple millimeter continuum sources, from where several molecular outflows emanate.
\end{abstract}



\section{Introduction} \label{sec:intro}
Carina, or NGC 3372, contains two very large and massive star clusters known as Trumpler 14 (Tr14) and Trumpler 16 (Tr16) that produce large amounts of ionizing light at UV wavelengths. These two massive clusters together contain dozens of O-type stars on the main sequence that reach bolometric luminosities surpassing the 10$^7$ L$_\odot$ \citep{s&b2008}. All this radiation permeates the vicinity of the nebula revealing the well-known irradiated dusty pillars, protoplanetary disks, and jets \citep{Reiter2013, Reiter2019, Reiter2020}. Therefore, this is an ideal nebula with extreme conditions to study the formation of new generations of stars in such harsh environments.

Recently, \citet{Cortes} presented a study using 1.3 mm continuum and C$^{18}$O(2$-$1), N$_2$D$^+$(3$-$2), $^{13}$CS(5$-$4), $^{12}$CO(2$-$1) spectral lines obtained from Atacama Large Millimeter/Sub-millimeter Array (ALMA) observations of the Carina Pillars HH 901 and 902. These observations revealed the outflows and the dusty compact disks that are exciting the extended and irradiated Herbig-Haro (HH) objects far from the pillars, and they estimated that the pillars would be photoevaporated in timescales between 10$^4$ and 10$^5$ yr by the massive stars mainly localized in Tr14 and Tr16. In addition, the circumstellar disks associated with the HH objects 901/902 would be likely exposed to the strong UV radiation soon after these scales of time, so they will transform into proplyds, like those observed in other star forming regions like in Orion \citep{odell1999, Henney1999}. In an even more recent study, \citet{Reiter2020} also presented sensitive and high angular resolution ALMA observations of the tadpole, a relatively small, and strongly irradiated globule inside the Carina Nebula which surrounds the HH 900 object. The ALMA observations obtained in \citet{Cortes} for the HH 901/902, revealed the disk and outflow system inside of the pillars. For this object, \citet{Reiter2020} reported a timescale for the photoevaporation of $\sim$ 4 $\times$ 10$^{6}$ yr, an order of magnitude larger than the dynamical age determined for HH 902 \citet{Cortes}, a pillar closer to the Trumpler clusters. 

\begin{figure*}[!]
\begin{center}
\includegraphics[scale=0.6]{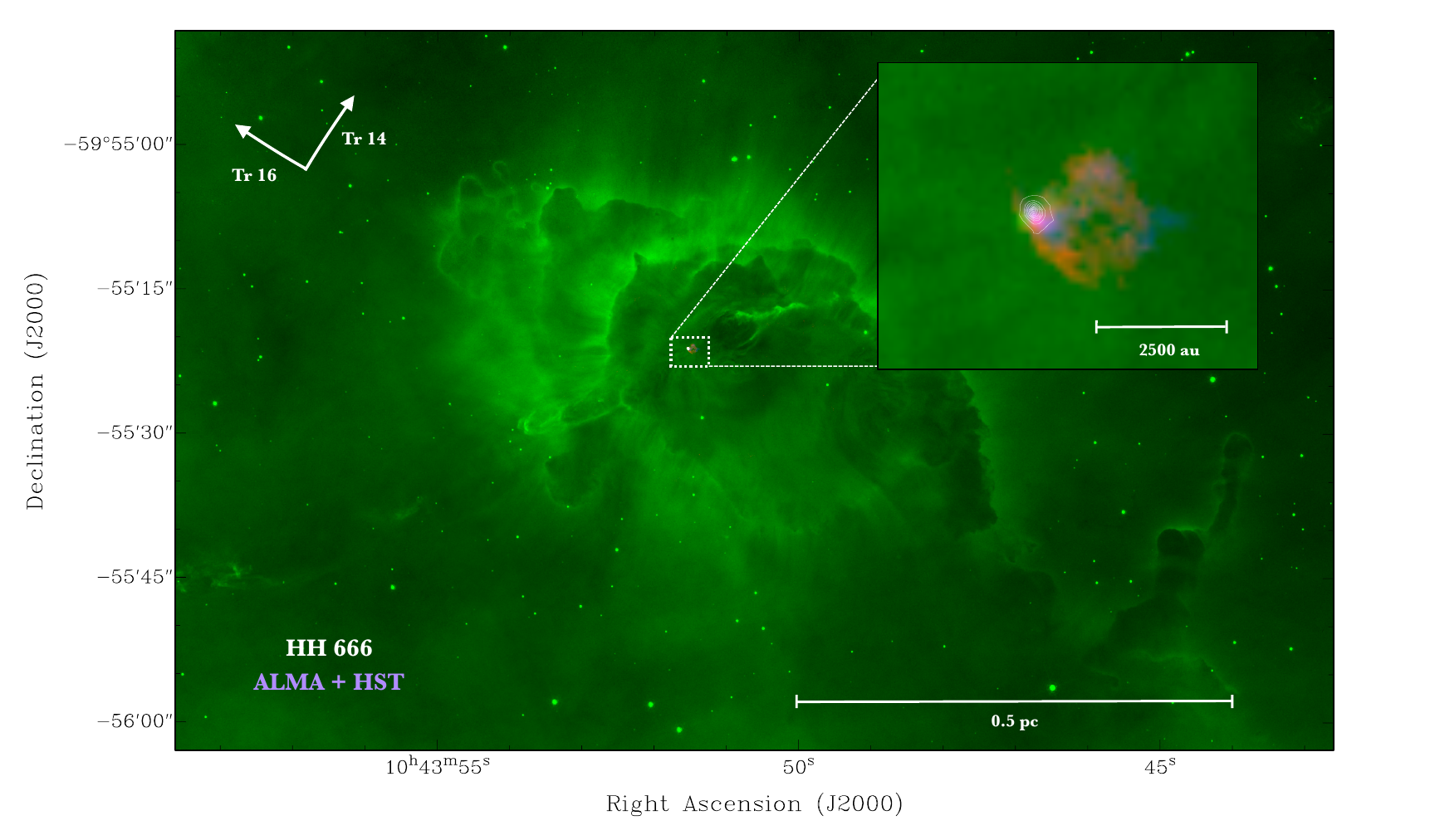}
\caption{A combined (optical$+$radio) HST and ALMA image is shown for the globule HH 666. The optical image, in green, is the H$\alpha$ emission and is showing extended optical jet, and obscured globule that houses the protostellar source. It also shows a millimeter continuum image at 1.3 mm frequency (white contours). The upper right panel shows the zoom image. The 1.3 mm continuum sources are denoted by white contours. The contour levels correspond to 10\% to 90\% of the peak flux with a step of 10\%. The blue and red colors show the moment zero images obtained from CO(2$-$1) emission with the velocity range of -10.19 to -10.83 km s$^{-1}$ for the redshifted emission, and from $-$11.46 to $-$18.45 km s$^{-1}$ for the blueshifted emission. The peak of the millimeter continuum emission is  5.87 mJy beam$^{-1}$ but there is still extended emission at lower levels. The local standard rest (LSR) system velocity of the globule is about $-$10 km s$^{-1}$. The white arrows indicate the directions where the star clusters Tr 14 and Tr 16 are located.}
\label{HH666_1}
\end{center}
\end{figure*}

Both ALMA studies are an important success in revealing the disk-outflow systems that are tremendously embedded in gas$+$dust contained in the globule HH 900 and pillars HH 901 and 902 that cannot be studied at optical or infrared wavelengths. Therefore, we present a complementary study using the ALMA that includes the pillars and globules containing the objects in the Carina Nebula: HH 666, HH 900, HH 1004, HH 1006, HH 1010, and HH 1066. This study allows us to reveal the exciting sources of the HH objects, estimate with more accuracy their physical conditions, and study how such harsh environments affect their evolution, and, in particular, constrain the formation of planets inside of the circumstellar disks.   \\

\begin{table*}[!]
\resizebox{\textwidth}{!}{
\begin{tabular}{l*{9}{c}r}
\multicolumn{10}{c}{}&\\
\hline
\hline
\multicolumn{2}{c}{Source}&
\multicolumn{3}{c}{Position}&
\multicolumn{2}{c}{Deconvolved Size}&
\multicolumn{2}{c}{Flux}&
\multicolumn{1}{c}{RMS\footnote{The theoretical rms for all sources is 0.046 mJy.}}&\\
\hline
&{} & $\alpha_{2000}$ &{$\delta_{2000}$} & {$\theta_{maj}$} &{$\theta_{min}$} &P.A.&{Peak} &{Integrated}& {Observed}\\
&{} & [h:m:s] & [$^{\circ }$ $^{\prime }$ $^{\prime\prime }$ ] & [milliarcsecond] & [milliarcsecond] &[$^{\circ }$] & {[mJy beam$^{-1}$]} &{[mJy]}& {[mJy]}\\
\hline
\hline
HH 666& & 10:43:51.532 & -59:55:21.18 & 85.2 $\pm$ 4.6 & 68.60 $\pm$ 3.5 &31 $\pm$ 12& 5.87 $\pm$ 0.17 & 14.94 $\pm$ 0.57 & 0.071 & \\
HH 900 A& & 10:45:19.301 & -59:44:22.53 & 576 $\pm$ 52 & 396 $\pm$ 42 &155 $\pm$ 12 & 2.41 $\pm$ 0.14 & 8.37 $\pm$ 0.61 & 0.043 &  \\
HH 900 B& & 10:45:18.797 & -59:44:23.79 & 272 $\pm$ 52 & 109 $\pm$ 70 &62 $\pm$ 17& 0.75 $\pm$ 0.04 & 1.09 $\pm$ 0.09 & 0.043 & \\
HH 1004 A& & 10:46:44.989 & -60:10:14.73 & 101 $\pm$ 9 & 62 $\pm$ 17.5 &153 $\pm$ 13 & 3.51 $\pm$ 0.13 & 7.42 $\pm$ 0.38 & 0.053 & \\
HH 1004 B& & 10:46:44.808 & -60:10:21.20 & 61 $\pm$ 30 & 44 $\pm$ 22 &177 $\pm$ 71 & 0.62 $\pm$ 0.06 & 0.90 $\pm$ 0.15 & 0.053 & \\
HH 1004 C& & 10:46:44.969 & -60:10:15.12 & 36 $\pm$ 14 & 17 $\pm$ 18 &178 $\pm$ 49 & 1.32 $\pm$ 0.04 & 1.49 $\pm$ 0.09 & 0.053 & \\
HH 1006&  & 10:46:32.902 & -60:03:53.84 & 1139 $\pm$ 89 & 945 $\pm$ 75  &43 $\pm$ 17 & 0.38 $\pm$ 0.02 & 9.41$\pm$ 0.73 &  0.050  & \\
HH 1010& & 10:41:48.640 & -59:43:38.04 & point source & point source &$-$ &  3.96 $\pm$ 0.09  & 4.12 $\pm$ 0.17 & 0.097  & \\
HH 1066& & 10:44:05.395 & -59:29:40.24 & 373 $\pm$ 85 & 211 $\pm$ 92 &  170 $\pm$ 26 & 1.37 $\pm$ 0.18 & 3.86 $\pm$ 0.66 &   0.053 & \\
\hline
\hline
\end{tabular}}
\caption{Physical parameters of the protoplanetary disks $+$ envelopes detected in Carina from 1.3 mm continuum images for different sources using the reduced ALMA data.}
\label{table1}
\end{table*}

\section{Observations and data reduction} \label{sec:obs}

The observations of the pillars and protostellar objects (HH 666, HH 900, HH 1004, HH 1006, HH 1010, and HH 1066) were performed with the Atacama Large Millimeter/sub-millimeter Array (ALMA) at Band 6 in 2018 January 1 and 4 (C43-6), and April 20 (C43-3) as part of the Cycle 5 program 2017.1.00912.S. Forty three antennas with diameters of 12 m were used for the observations, yielding baselines from 15 $-$ 2516 m (11.5 $-$ 1935 k$\lambda$) for C43-6 configuration and 15 $-$ 500 m (11.5 $-$ 384.6 k$\lambda$) for C43-3 configuration. The primary beam at this frequency has a full width at half-maximum (FWHM) of about 25$^{\prime\prime}$, covering well the bulk of molecular and dusty emission from the pillars.

The 1.3 mm continuum and molecular line images were obtained from the two configurations, C43-6 and C43-3, that had source integration times of 9 and 3.5 min, respectively. We used four spw that were centered at the frequencies, 218.014 GHz(spw0) for 1.3 mm continuum and 219.578 GHz(spw1), 231.239 GHz(spw2), and 230.556 GHz(spw3) for line analysis. The channel widths were 21.486 km s$^{-1}$ for the continuum and 333.328, 633.038 and 634.912 m s$^{-1}$ for the C$^{18}$O, N$_2$D$^+$ and CO respectively. The total bandwidth for the continuum is about 4.3 GHz. Three molecular lines of C$^{18}$O(2$-$1) ($\nu_\mathrm{rest}$= 219.56035 GHz), N$_2$D$^+$(3$-$2) ($\nu_\mathrm{rest}$=231.31990) and $^{12}$CO(2$-$1) ($\nu_\mathrm{rest}$=230.53800 GHz) were observed in spw1, spw2 and spw3, respectively. System temperatures were between 100 to 150 K and precipitable water vapor was between 1.3 mm and 2.2 mm. The ALMA calibration included simultaneous observations of the 183 GHz water line with water vapor radiometers, used to reduce atmospheric phase fluctuations. Quasars, J0904-5735, and J1107-4449 were used for the band-pass and flux calibrations. J1032-5917 was used for correcting the gain fluctuations.

The data reduction was done using the Common Astronomy Software Applications \citep[CASA,][]{CASA} version 5.1. We concatenated the data from both configurations (C43-6 and C43-3) with the \texttt{UVCONCAT} task. Imaging of the calibrated visibilities was done using the \texttt{TCLEAN} task. We used the briggs weighting (robust = 0.5) in the \texttt{TCLEAN} task to obtain the continuum and spectral lines images. We report the parameters obtained from a two-dimensional Gaussian fit in CASA to the continuum images at 1.3 mm, using the gaussfit tool (within the CASA \texttt{viewer}), which subtracts the sky component to a selected region for the emission distribution. We obtained the deconvolved sizes from the synthesized beam of each image and this Gaussian fit. The results of the physical parameters (position, deconvolved size, flux, and rms) are shown in Table \ref{table1}. We attempted self-calibration on the continuum emission, but unfortunately, the resulting maps did not improve significantly. Thus, we stick with the non-self-calibrated maps.


 \section{Results and Discussion} \label{sec:res}
 
\begin{figure*}[!]
\begin{center}
    \includegraphics[width=0.9\textwidth]{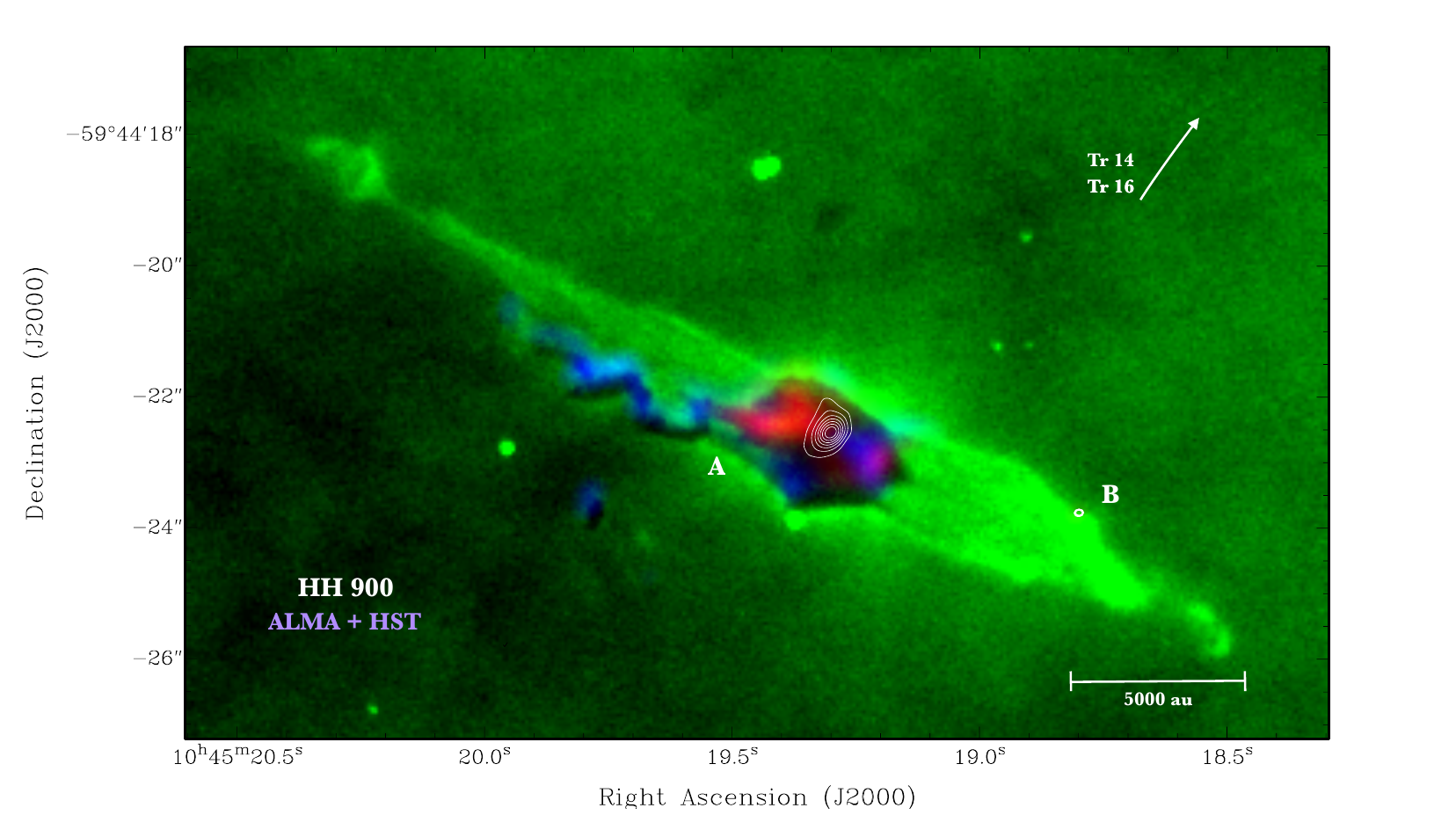}
    \caption{A combined HST (optical) $+$ ALMA (radio) image is shown for the globule HH 900. The optical image (green) is the H$\alpha$ emission about the optical jet and obscured globule that houses the protostellar source evidenced by 1.3 mm emission (white contours varying from 30$\%$ to 90$\%$ of the peak emission, in steps of 10\%). The peak of the millimeter continuum emission is 2.41 mJy beam$^{-1}$ for component A and 0.75 mJy beam$^{-1}$ for component B. Also shown in blue and red colors is the CO(2$-$1) moment zero for the molecular outflow associated with the protostar (component A). Component B does not have any associated molecular outflow. In this case, the blue color corresponds to the blueshift molecular gas, and the red color corresponds to the redshift molecular gas. The LSR system velocity of the globule is about $-$33.72 km s$^{-1}$. The white arrow indicates the direction where the star clusters Tr 14 and Tr 16 are located.  
    }
\label{HH900_1}
\end{center}
\end{figure*}

 Our data reveal for the first time the bipolar molecular outflows emerging from HH 666/1004/1006/1010 and 1066 YSOs (Young Stellar Objects) which corresponds very well to the optical jets previously studied by Hubble Space Telescope (HST) using the Advanced Camera for Surveys (ACS) (see Figures \ref{HH666_1}, \ref{HH900_1}, \ref{hh1004_1}, \ref{HH1006_1}, \ref{hh1010_f1} and \ref{HH1066_1}). In the following subsections, we show the analysis for each separate object. We show their gas and dust structures obtained with the ALMA observations together with some estimated physical values, such as fluxes, masses, sizes, and the comparison with their respective optical and infrared bulk emissions. The parameters obtained from the observations and the reduction of the data carried out with CASA are shown in Table \ref{table1}.
 
\subsection{HH 666}  

The HH 666 object, known as the axis of evil, was first reported by \citet{smith2004} using ground-based optical and near-IR (infrared) observations that revealed the bipolar flow emerging from the head of a dusty globule pointing in the direction of $\eta$ Carinae and Tr 16.  A protostar (listed as Class I object) embedded within the dense globule is proposed as the possible driving source of the optical jet (detected at the center of the globule at the position in the sky of $\alpha = 10^{h}43^m 51^s.3$, $\delta = - 59^{\circ} 55^{\prime} 21^{\prime \prime}.2$, J2000), see \citet{smith2004}. In a later study, \citet{smith2010} reported new H$\alpha$ observations from the HST, corroborating the jet structure associated with the HH object, the extension of jet component ($\sim$4$^{\prime}$.5 or $\sim$3 pc), and the detection of a weak object located at the same position as the previous study. This object was also detected in Spitzer near-IR observations \citep{smith2010} 

In recent work, \cite{Reiter2015a} presented near-IR spectroscopic studies of [Fe II] obtained with the Folded-Port InfraRed Echelette (FIRE) spectrograph on the Baade/Magellan Telescope. These observations reveal that  [Fe II] emission is tracing a highly collimated protostellar jet embedded in the dense pillar of gas and dust with an extension of about 40 arcsecs. They also found a clear connection between the previously revealed jet driving source and the extent of the collimated jet.  Finally, their studies of [Fe II] and H$\alpha$ showed differences, especially in areas very close to the protostellar source, where the images in H$\alpha$ did not reveal any emission. 

In Figure \ref{HH666_1}, we present the optical image of the {\it HST/ACS} presented in \citet{smith2010} overlaid with our resulting ALMA 1.3 mm continuum and line images of the globule HH 666. The optical image (green) corresponds to H$\alpha$ emission and is overlapped with the ALMA high velocity CO(2$-$1) molecular emission (blue and red) and the 1.3 mm continuum emission (white contours). The CO emission is tracing the bipolar outflow very close to the exciting source, which moves at an LSR system velocity of $-$10 km s$^{-1}$. The angular size is 1.6$^{\prime\prime}$ which at 2.3 kpc represents a size of 3680 au for HH 666.

The 1.3 mm continuum emission is only located in the axis of the optical jet detected in H$\alpha$ at $\alpha = 10^{h}43^m 51^s.5$, $\delta = - 59^{\circ} 55^{\prime} 21^{\prime \prime }.1$ (J2000), perfectly matching the near-IR source detected by \citep{smith2004} (see Figure \ref{HH666_1}) and reveals a compact source corresponding to an embedded structure in the central zone of the globule. We think that this object is composed of a dusty envelope and circumstellar disk corresponding to the driving source of the protostellar optical jet, and for this millimeter continuum emission, we obtained an integrated flux of 14.94 $\pm$ 0.57 mJy and an observed rms noise of 0.071 mJy (see Table \ref{table1}) for the continuum image, which is very close to the theoretical noise value for this configuration (43 antennas), frequency (230 GHz), bandwidth (4.3 GHz) and integration time (12.5 min.) and is 0.046 mJy. From a Gaussian fit in CASA for the continuum source and the synthesized beam size (0$^{\prime\prime}$.07 $\times$ 0$^{\prime\prime}$.05), we obtained a deconvolved size of 0$^{\prime\prime}$.0852 $\pm$ 0$^{\prime\prime}$.0046 $\times$ 0$^{\prime\prime}$.0686 $\pm$ 0$^{\prime\prime}$.0035 with an PA of 31$^{\circ}$ $\pm$ 12$^{\circ}$, that corresponds to an approximate physical size of 200 au $\times$ 150 au, which is according to the physical sizes of the protoplanetary disks ($\sim$100 au).

From this detected dust emission, we can estimate the total mass for these structures, even for the case of component B, which is much fainter than component A. At millimeter wavelengths, the emission can be assumed to be optically thin, and we can then estimate the dust mass from the following expression:

\begin{equation}
M_d=\frac{D^2 \, S_\nu}{\kappa_\nu \, B_\nu(T_d)}
\label{eq1}
\end{equation}

\noindent
where S$_{\nu}$ is the flux density, D is the distance to the considered sources, we assume that it is the same as the Carina Nebula  \citep[2.3$\pm$0.1 kpc,][]{s&b2008}, $\kappa_\nu$= 0.015 cm$^2$ g$^{-1}$ is the dust mass opacity for a dust-to-gas ratio of 100 appropriate for 1.3 mm \citep{ossenkopf1994} and B$_\nu(T_d)$ is the Planck function for the dust temperature T$_d$. Given the observing frequency of our data ($\sim$ 225 GHz), we considered the Rayleigh-Jeans  regime. We compute the values for the masses of this (and the others) source and show them in Table \ref{table2} from the measured fluxes presented in Table \ref{table1}. For this source, we found that the dust mass associated with the detected continuum structure ($\sim$ 15 mJy) is between 0.3$-$0.7 M${\odot}$, assuming temperature values between 20 K and 50 K as \citet{Cortes}.   

\begin{table}[!]
\centering
\begin{tabular}{l*{7}{c}r}
\hline
\hline
\multicolumn{1}{c}{Source}&
\multicolumn{4}{c}{\,\,\,\,\,Disk masses\footnote{We estimate different mass values by varying the assumed temperature to compare the mass variation.}}&
\multicolumn{2}{c}{\,\,\,\,\, RMS masses\footnote{Masses estimated from image noise at 4$\sigma$.}}& \\
&{} & 20 K &40 K &50 K  &   & \\
\cline{3-5}
&{} && [M$_\odot$]&  &  &[M$_{Jup}$] & \\
\hline
\hline
HH 666 && 0.77&0.38&0.30&&6.14 \\
HH 900 A&& 0.43& 0.21& 0.17 & &3.72\\
HH 900 B&&0.05 & 0.02& 0.02 && 3.72\\
HH 1004 A&&0.38 & 0.19& 0.15 && 4.58\\
HH 1004 B&&0.04& 0.02& 0.01 &&4.58\\
HH 1004 C&&0.07& 0.03& 0.03 &&4.58\\
HH 1006  &&0.48&0.24 & 0.19 && 4.32\\
HH 1010 && 0.21& 0.10& 0.08 && 8.39\\
HH 1066  &&0.19& 0.09&  0.07 && 4.58\\
\hline
\hline
\end{tabular}
\caption{Dust masses obtained by the emission detected at 1.3 mm for the compact systems (discs+envelopes) associated to each HH object studied.}
\label{table2}
\end{table}

As mentioned earlier, we also show the molecular CO(2$-$1) emission corresponding to the outflow in the HH 666 object. In Figure \ref{HH666_1}, we can see the molecular emission plotted in blue and red colors corresponding to the blueshifted and redshifted gas, respectively. From these maps, we only detected CO(2$-$1) emission (blueshifted$+$redshifted) toward the western side of the ALMA 1.3 mm continuum source, revealing a monopolar outflow, which is associated with the extended HH 666 object. We did not detect molecular CO(2$-$1) emission associated with the large globule itself, maybe because its emission is either extended and complex, so our ALMA observations could not recover quite well the extension of the globule. Another possibility is that this is an old globule with a quite small quantity of molecular gas. \citet{klassen2019} using ALMA Compact Array (ACA) observations of pillar 3 (corresponding to the HH 666 globule) reported clumpy and dispersed CO(2$-$1) emission. Despite the larger beamsize of their observations with respect to ours (i.e. 6 arcsec or about 20 times larger), the extended emission of the globule was poorly recovered, we then favor the interpretation that our observations are losing most of the extended emission.

Since the detected CO emission is faint, it is difficult to say anything about the orientation, extension, or morphology of the molecular outflow. From Figure 1, this appears to have a West to East orientation with mostly the compact emission arising towards the western side. In addition, we cannot give an estimation for the mass of the globule, and its respective outflow due to not tracing their complete structure.     

\begin{table*}[!]
\centering
\small
\begin{tabular}{l*{13}{c}r}
\hline
\hline
\multicolumn{1}{c}{Source}&
\multicolumn{5}{c}{Pillar/Globule}&
\multicolumn{7}{c}{\hspace{1cm}Outflow\footnote{Red and blue superscript refer to detected redshifted and blueshifted gas.}}&\\
\cline{2-6} \cline{9-12}
&&I$_{\nu}$\footnote{Average intensity for the pillar or globule.}&d$v$\footnote{Velocity range for the pillar or globule flux.}&Size\footnote{We obtain the size by multiplying $\theta_{max}\times\theta_{min}$.}&Mass&{}&{}&I$_{\nu}^{\, red,}$\footnote{Average intensity for the outflow.} \hspace{0.3cm}I$_{\nu}^{\, blue}$&d$v^{red,}$\footnote{Velocity range for the outflow flux.}\hspace{0.3cm}d$v^{blue}$&Size$^{red}$ \hspace{0.3cm} Size$^{blue}$&Mass\footnote{It is the total mass of the outflow, that is, the sum of the mass of the redshifted and the blueshifted outflow.}&\\
&{}&[
$\frac{\rm{mJy}}{\rm{beam}}$
]&[km s$^{-1}$]&[arcsec$^2$]&[M$_\odot$]&{}&{}&[
$\frac{\rm{mJy}}{\rm{beam}}$
]&[km s$^{-1}$]&[arcsec$^2$]&[M$_\odot$]&\\
\hline
\hline
HH 900&&60&2.7&5.2&4$\times10^{-3}$&{}&{}&46.9 \, 23.5&3 \, 1&2 \, 2 & 1.4$\times10^{-3}$&\\
HH 1006&&50&3.2&11.4&7$\times10^{-3}$&{}&{}&44 \,\,\,\, 17.6&\,\,\,\,3 \, 2.5&3.2 \, 1.2& 2.1$\times10^{-3}$&\\
HH 1010&&50&2.2&170&7$\times10^{-2}$&{}&{}&\hspace{-.1cm}3.3 \,\,\, 4&4 \, 2& \,\, 2.5 \, 0.25& 1.5$\times10^{-3}$&\\
\hline
\hline
\end{tabular}
\caption{Gas masses obtained by the emission detected for the CO(2$-$1) emission at 230 GHz for the pillars/globules and outflows associated with each HH object studied. We estimate the mass using an excitation temperature of T$_{ex}$=50 K, an abundance ratio between the molecular hydrogen and CO of $X_\frac{H_2}{CO}\sim$10$^4$, and a distance to Carina nebula of 2.3 kpc. We only included the sources for which there was CO molecular detection.}
\label{gasmass}
\end{table*}

\subsection{HH 900}

HH 900 is a dark globule located very close to the central zone of the Carina Nebula,  a region strongly affected by the extreme UV radiation generated by nearby massive stars to the star cluster Trumpler 16, and the massive and evolved variable star $\eta$ Carinae located about $\sim$3 pc from the HH 900. It has an angular size of the order of 1$^{\prime\prime}$ (equivalent to a spatial scale of $\sim$0.01 pc). This source is also known as the tadpole due to its unusual morphology that simulates a baby frog, made up of a dense globule (tadpole head) followed by an elongated and compact tail that extends in the same direction as the axis of the optical jet revealed by the images of the \textit{HST} (see Figure \ref{HH900_1}). Using these observations, \citet{smith2010} studied the components of the bipolar structure of the optical jet, and globule traced by the gas emission of H$\alpha$ corresponding to externally ionized gas. The asymmetric morphology revealed at optical wavelengths has not been explained since it differs from the typical jet morphologies that present a collimated and very symmetrical bipolar structure \citep{Reipurth2001,Reipurth2019,Noriega-Crespo}. The extreme UV radiation field is illuminating throughout the length of the protostellar jet with a projection close to the plane of the sky and an inclination angle for the jet $\sim$ 10$^{\circ}$ \citep{Reiter2015a_hh900}.\\ 

They studied the high-velocity component of the protostellar jet through near-infrared (NIR) observations for the line of [Fe II] and that, unlike the observations of H$\alpha$, showed an external collimated jet structure that is extending from east to west of the globule, presenting a remarkable symmetry that argues for the fact that the driving source would have to be located inside the center of the globule although they did not find IR emission from the protostar inside the globule, suggesting the fact that the protostar may be embedded inside remaining invisible due to the high column density. \citet{shinn2013} detected two sources associated with the HH 900 object through observations of [Fe II] at 1.64 $\mu$m but until now, no study could corroborate that these detected sources were generating the jets because no direct connection was found between the extended flows and the compact sources detected by this study (YSO also cataloged by \textit{Spitzer}, \citeauthor{povich2011}, \citeyear{povich2011}).

Subsequently, \citet{Reiter2019} studied the HH 900 source at optical wavelengths with \textit{MUSE}/\textit{VLT} observations detecting several emission lines (hydrogen recombination lines and some forbidden emission lines) that also trace the outflow. Their observations were also unable to detect the protostar embedded within the tadpole, although the emission was detected for different lines that are tracing the collimated bipolar jet at different densities (for example [Fe II], [Ca II], and [Ni II]), is emanating from the globule, which reinforces the idea of that the protostellar source must be housed inside it.

More recently \citet{Reiter2020} reported using ALMA observations the bipolar molecular outflow traced by the CO emission arising from the protostar called HH 900 YSO. The outflow is bi-conical and matches very well the overall morphology of the optical jet. The estimated mass for the total globule is about 2 M$_\odot$ (from the dust continuum), which results in a lifetime of 4 Myrs of \citet{Reiter2020}, with the assumption of a constant photo-evaporation rate. 

Our results obtained from the ALMA observations for the continuum emission at 1.3 mm reveals two compact sources called A and B (white contours in Figure \ref{HH900_1}). These two compact objects were already reported in previous ALMA observations presented in \citet{Reiter2020}. The millimeter emission is shown in Figure \ref{HH900_1} as white contours and is likely associated with the dust emission of the circumstellar envelope and flatted disk (see for example \citet{Reiter2020, Reiter2020b}). We note that the emission is located at the center of the tadpole's head from which emerges the bipolar molecular outflow also detected thanks to the high angular resolution of ALMA \citet{Reiter2020}. Our observations also revealed a second component traced by the continuum emission (component B) located in the southeastern part of the optical jet in the area strongly illuminated by external ultraviolet radiation. From the Figure, we can see that there is no molecular outflow associated with this component, maybe because this source is prestellar, as already discussed in \citet{Reiter2020}. 
\citet{Reiter2019} also rejected the premise that this detected object (named PCYC 838) was driving a microjet. They dismissed the idea by arguing that the symmetric [Fe II] jet is centered on the globule and not the star, in addition to the lack of evidence of different velocity components in the reported spectra, concluded that a likely explanation is that this object is coincidentally aligned with the observed jet.

From the continuum emission, we obtained an integrated flux of 8.37 $\pm$ 0.61 mJy and 1.09 $\pm$ 0.09 mJy for components A and B respectively, and an observed rms noise of 0.043 mJy (see Table \ref{table1}). From the synthesized beam size (0$^{\prime\prime}$.34 $\times$ 0$^{\prime\prime}$.27), we obtained a deconvolved size for component A of 0$^{\prime\prime}$.576 $\pm$ 0$^{\prime\prime}$.052 $\times$ 0$^{\prime\prime}$.396 $\pm$ 0$^{\prime\prime}$.042 with an PA of 155$^{\circ}$ $\pm$ 12$^{\circ}$, with an approximate physical size of 1330 $\times$ 910 au.

From the millimeter continuum emission, we estimate the dust mass associated with the structure detected by our observations, following the equation described in the previous subsection (\ref{eq1}), giving values between 0.17$-$0.43 M$_\odot$, assuming temperatures between 20 K and 50 K for the component A, which is the component associated with the detected CO(2-1) bipolar molecular outflow shown in red and blue colors in the Figure \ref{HH900_1}. This line emission at 230 GHz is tracing perfectly the tail of the tadpole (blueshifted CO gas) extending along the axis of the optical jet, just like its counterpart (redshifted CO gas). We can deduce that the protostar is a Class 0/I object from the mass-loss rate ($\sim$ 5$\times$ 10$^{-6}$ M$\odot$ yr$^{-1}$) reported by \citet{Reiter2015a_hh900}. The continuum emission of this source possibly includes the disk and envelope, which reveals the emission from the disk and the envelope around the driving source. For component B we estimated a dust mass between 0.02$-$0.05 M$_\odot$ under the same assumptions.

We note in Figure \ref{HH900_1} that the molecular outflow is aligned to the optical jet, and expands at velocities between  $-$32 and $-$36 km s$^{-1}$. The flux detected in CO is tracing the innermost parts of the dark globule connected to the extent of the outflow with the driving source detected by the continuum emission. We integrate LSR velocities from $-$29.27 to $-$33.08 km s$^{-1}$ for the redshifted emission oriented to the northwest, and from $-$34.99 to $-$40.70 km s$^{-1}$ for the blueshifted emission oriented to the southeast, for the CO outflow in the HH 900 object. The molecular outflow revealed by the CO is only evident for one component denominated component A.

\begin{figure}[!]
\centering
    \includegraphics[width=0.5\textwidth]{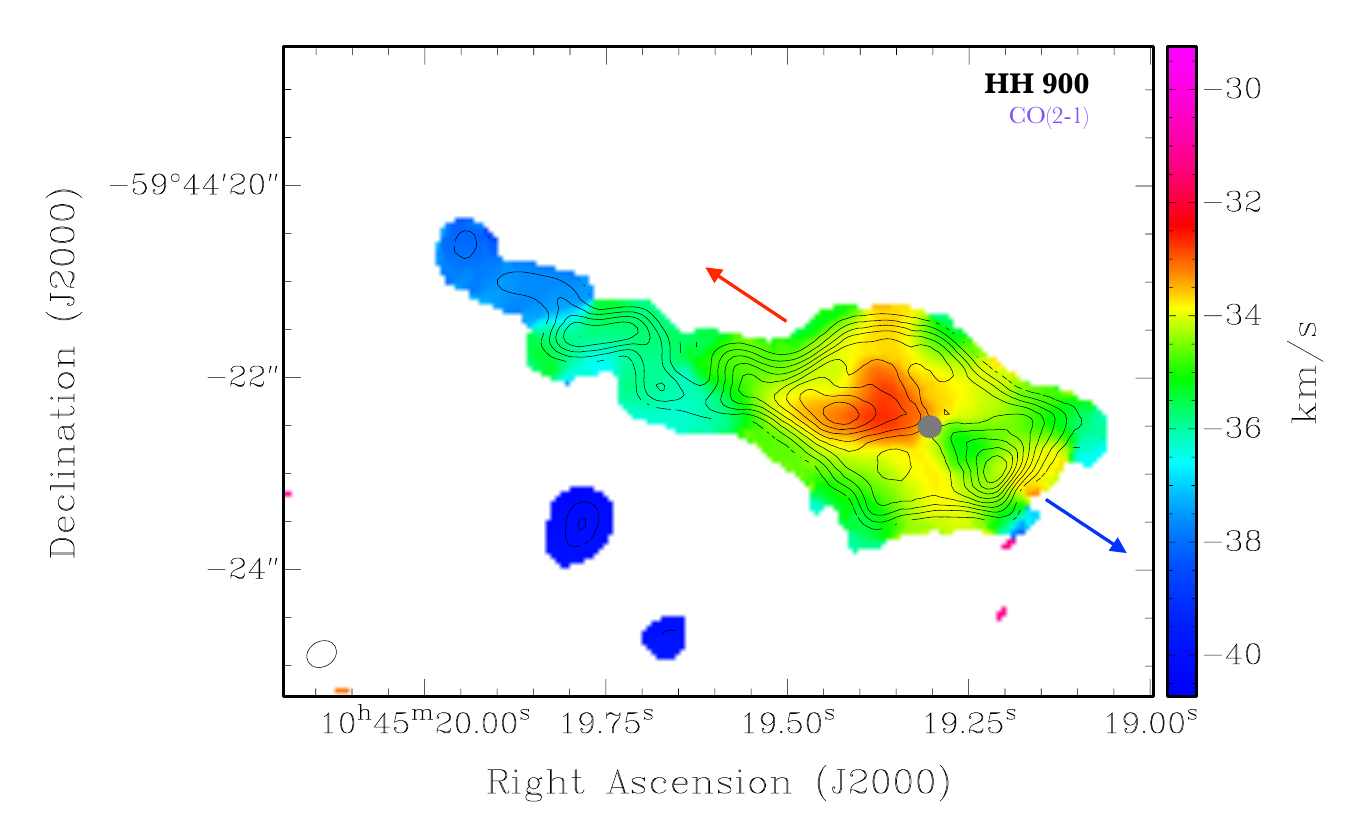}
    \includegraphics[width=0.515\textwidth]{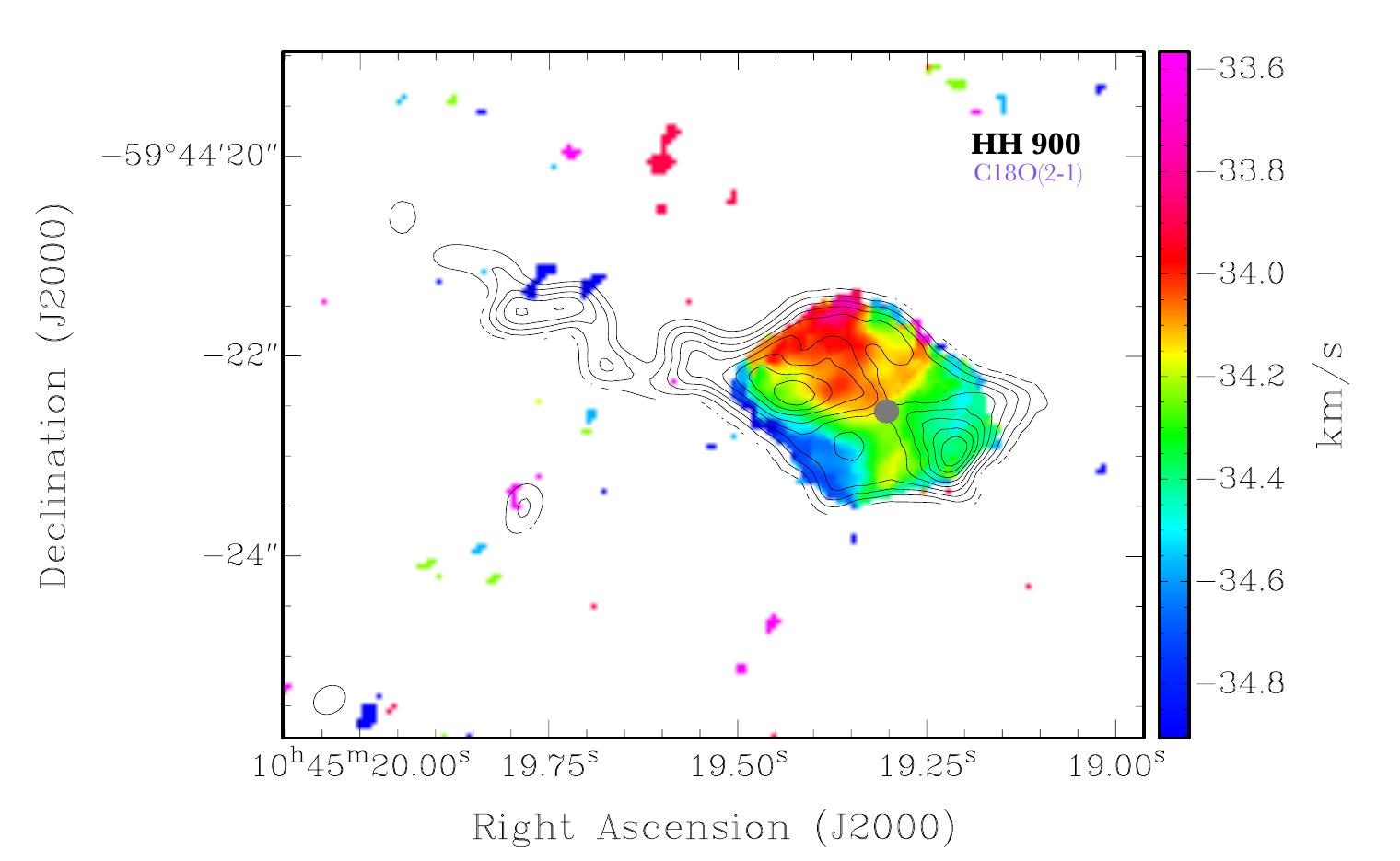}
    \caption{The upper panel shows the moment zero (contours) and moment one (colors) for CO(2$-$1), and the lower panel shows the moment zero (contours) of CO and moment one (colors) of C$^{18}$O (2$-$1) of the globule HH 900. The contours vary in the range from 40\% to 90\% of the peak emission, in steps of 12\%. The peak of the CO (2$-$1) emission is 0.33 Jy Beam$^{-1}$ km s$^{-1}$. The arrows (blue and red) show the direction of molecular outflow expansion. The gray circles in both figures represent the location of the compact source (system disk$+$envelope) detected in the continuum at 1.3 mm. The LSR radial-velocity scale bar is shown on the right.
    }
 \label{HH900_2}
\end{figure}

In Figure \ref{HH900_2}, we show the emission of molecular gas for the CO(2$-$1) (top panel) and C$^{18}$O(2$-$1) (bottom panel) detected lines. The moment zero of CO is shown in black contours for both panels of Figure \ref{HH900_2}, as a comparison between the emission of the molecules. The moment one is shown in color scale for CO (top panel) and for C$^{18}$O (bottom panel). 
It is observed that CO is perfectly tracing the globule, including the tail of the globule, hosting the compact source detected by the continuum (gray circle) and corresponding to the bipolar molecular outflow of CO also observed. The bipolar molecular outflow is also evident in these maps for both panels, with a northwestern orientation to the redshifted emission and a southeastern orientation to the blueshifted emission. If we analyze the velocities for the molecular outflow in this map, we note that the width velocity is $\sim$ 5 km s$^{-1}$, which suggests that the molecular flow is not a high velocity one.

The outflow extends no further than the emission detected for the rest of the HH 900 globule (bounded by the moment zero of CO). The different velocity components detected vary between the LSR range of $-$29.27 to $-$40.70 km s$^{-1}$, velocities that were used to integrate the moment zero (contours) and one (colors) of the CO emission. The average LSR velocity of the cloud is $\sim$ $-$34 km s$^{-1}$. The CO molecule best traces the globule as a whole, while the C$^{18}$O is evidencing more internal zones of the cloud where the protostar and envelope are localized, and the density is higher. This image (bottom panel) was obtained by integrating over the LSR velocity range of $-$33.23 to $-$35.23 km s$^{-1}$. The system LSR velocity of the cloud is $\sim$$-$33.90 km s$^{-1}$. For this case, the molecular emission shows a width of velocities $\sim$ 1 km s$^{-1}$, which confirms the result obtained with the CO map showing that these flows are not high speed.

 
 


We compute the gas mass using the observed CO transition J=2\,$\rightarrow$\ 1 from the following equation:

\begin{equation}
\begin{split}
\left[\frac{M_{H_2}}{M_\odot}\right]=1.2 \times10^{-15}\,T_\mathrm{ex}\,e^{\frac{16.59}{T_\mathrm{ex}}}
\,X_\frac{H_2}{CO} 
\left[\frac{\int \mathrm{I_\nu dv}}{\mathrm{Jy\,km\,s}^{-1}}\right] \\
\quad
\left[\frac{\theta_\mathrm{maj}\,\theta_\mathrm{min}}{\mathrm{arcsec}^2}\right]
\left[\frac{D}{\mathrm{pc}}\right]^2, 
\label{eq2}
\end{split}
\end{equation}

\noindent
where $X_\frac{H_2}{CO}\sim$10$^4$ \citep{scovi1986} is the abundance ratio between the molecular hydrogen (H$_2$) and carbon monoxide (CO), I$_{\nu}$ is the average intensity ($I_\nu$) for the pillar-globule/outflow, $dv$ is the velocity range for the pillar-globule/outflow, $D$ is the distance to the source of 2.3$\pm$0.1 kpc and T$_{ex}$ is the excitation temperature taken as 50~K, this is the highest temperature in the range used previously for the dust temperature, and it provides an upper limit for our mass estimation.

We obtained a value for the gas mass of the tadpole globule of 4 $\times$ 10$^{-3}$ M$_{\odot}$ using the emission lower than $\pm$1 km s$^{-1}$ from the system velocity, and the mass of the outflow was 1.4 $\times$ 10$^{-3}$ M$_{\odot}$ for velocities larger than $\pm$1 km s$^{-1}$ (see table \ref{gasmass}). The estimated values for both masses should be considered as a lower mass limit because the medium could be optically thick.

\begin{figure*}[!]
\centering
\includegraphics[width=2.2\columnwidth]{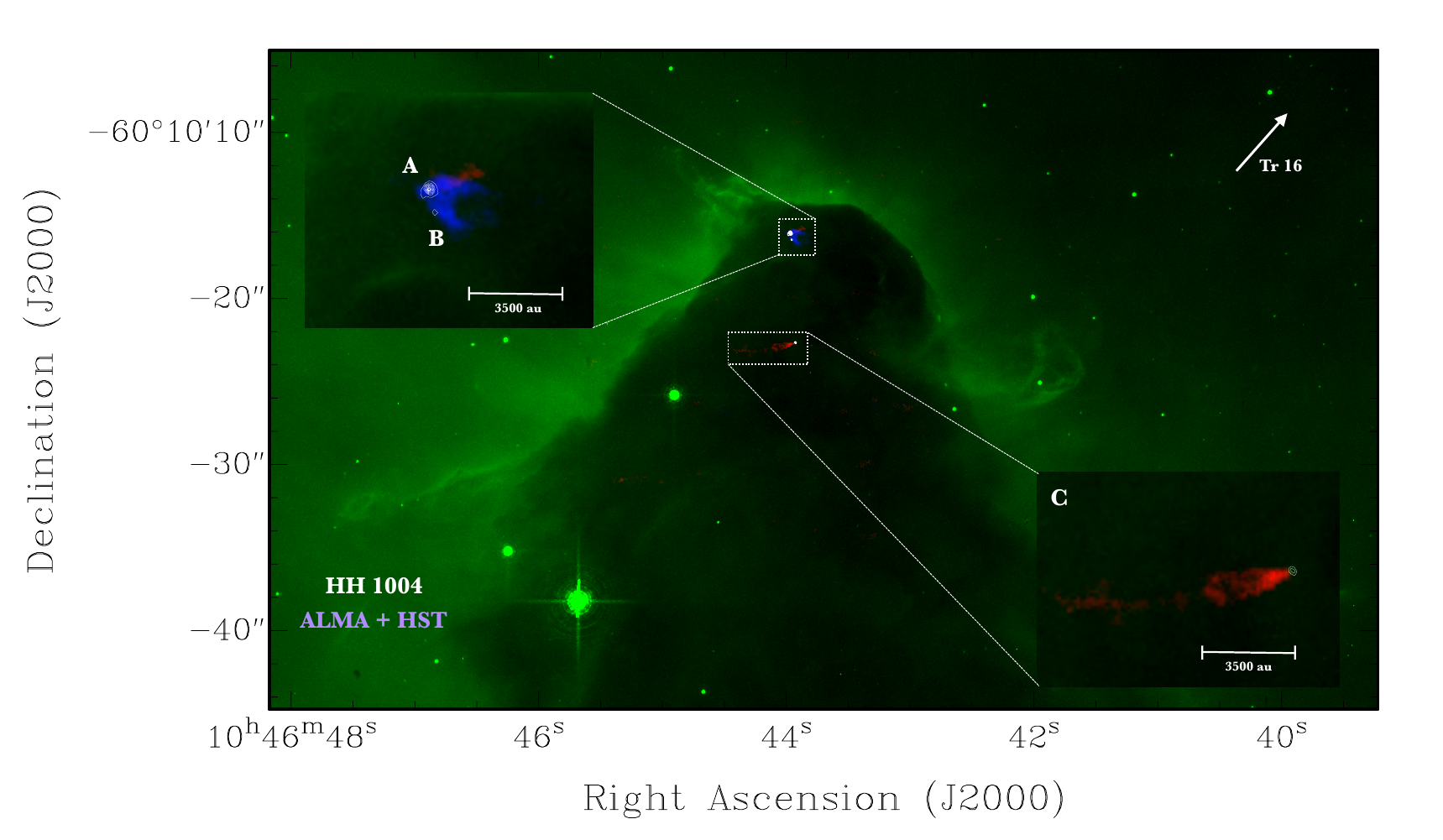}
    \caption{A combined (optical$+$radio) HST and ALMA image is shown for the pillar HH 1004. The optical image, in green, is the H$\alpha$ emission and is showing the optical jet and the pillar in which the compact source detected in the continuum at 1.3 mm is embedded and shown as white contours (varying from 15$\%$ to 90$\%$ of the peak emission, in steps of 10\%). Three compact sources were detected (components A, B, and C). The peak of the millimeter continuum emission is  3.51 mJy Beam$^{-1}$ but there is still extended emission at lower levels. Also shown in blue and red colors is the CO(2$-$1) moment zero for the molecular outflows associated with the compact sources. In this case, the blue color corresponds to the blueshift molecular gas and the red color corresponds to the redshift molecular gas. The LSR system velocity of the globule is about $-$20 km s$^{-1}$. The white arrow indicates the direction where the star cluster Tr 16 is located.
\label{hh1004_1}}
\end{figure*}

We calculate the theoretical value of the mass photo-evaporation rate ($\dot{M}$) for the globule HH 900 associated with the driving source following \citet{Cortes} (see their equation 3). The size of the globule was calculated using a Gaussian fitting in CASA, assuming a distance of 2.3 kpc. In this case, the globule is affected by ultraviolet radiation originating from the Tr 16 with a UV luminosity of (Q$_H$) = 9$\times10^{50}$ photons s$^{-1}$ \citep{s&b2008}. This luminosity (Q$_H$) translates into a flux F$_{EUV}=8.35\times10^{11}$ photons s$^{-1}$ cm$^{-2}$ assuming that the distance from the HH 900 globule to the cluster is $\sim$ 3pc. We obtained a mass photo-evaporation rate of 2.6$\times 10^{-5}$ M$_{\odot}$ yr$^{-1}$. This estimated value is consistent with the methodology described by \citet{Reiter2015a_hh900} (7 $\times$ 10$^{-6}$ M$_{\odot}$ yr$^{-1}$), taking into account that we have used the gas mass of the globule (instead of the dust mass) which translates into a difference of at least an order of magnitude in mass and a factor of $\sim$4 in size, so our value differs by an order of magnitude than the one presented by them. In this calculation we do not consider the optical depth value and since the globule is very optically thick that may account for the difference. We can estimate the photo-evaporation timescale of the globule HH 900 by dividing the estimated gas mass for the globule (4 $\times$ 10$^{-3}$ M$_{\odot}$) by this mass photo-evaporation rate, giving $\sim$ 150 yr, which will be the time in which the globule HH 900 will be photo-evaporated by the massive stellar cluster Tr 16 (see Table \ref{table4}). Using the dust mass we estimated for this globule (0.21 M$_{\odot}$), that corresponds to the whole continuum emission at 1.3 mm, the photo-evaporation timescale would be $\sim$8000 yr, $\sim$two orders magnitude less than the reported value reported by \citet{Reiter2020} (4 Myr), although it should be noted that our estimated dust mass differs by a factor of $\sim$10 and the photo-evaporation rate by a factor of $\sim$100.

\begin{figure*}[!]
\begin{center}
    \includegraphics[width=1.0\textwidth]{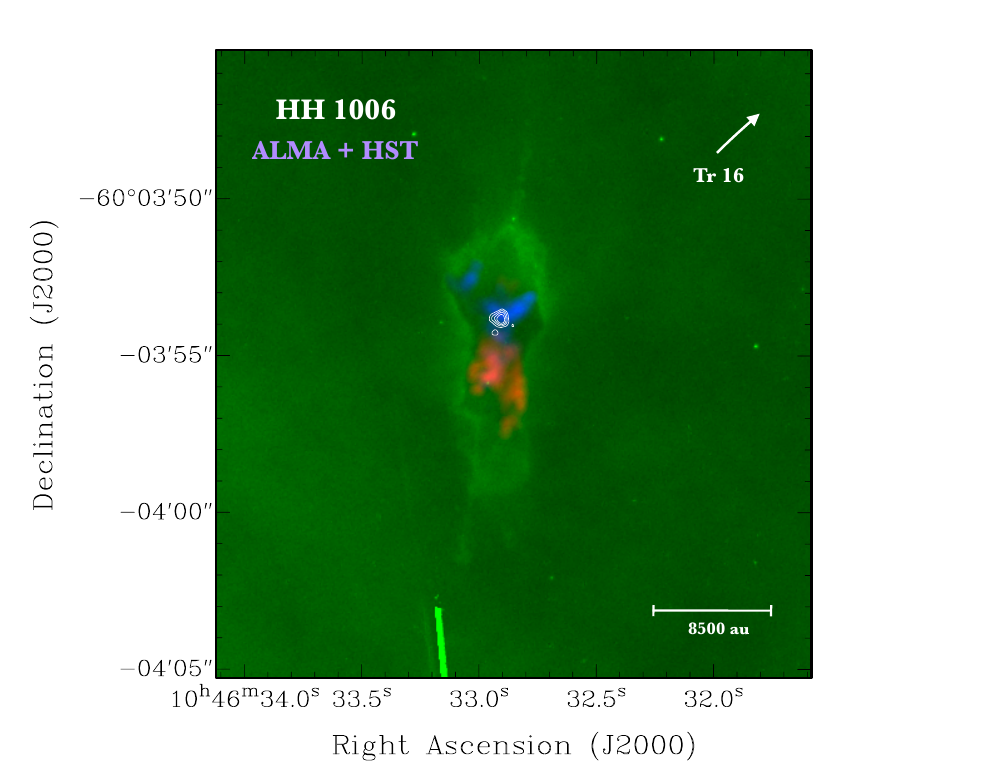}
    \caption{The figure shows a combined (optical$+$radio) HST and ALMA image for the globule HH 1006. The optical image (green) is the H$\alpha$ emission and is showing optical jet and the globule in which the compact source detected in the continuum at 1.3 mm is embedded, and shown as white contours (varying from 55$\%$ to 90$\%$ of the peak emission, in steps of 10\%). The peak of the millimeter continuum emission is  7.83 mJy beam$^{-1}$. Also shown in blue and red colors the CO(2$-$1) moment zero for the molecular outflows associated with the compact source. In this case, the blue color corresponds to the blueshifted molecular gas and the red color corresponds to the redshifted molecular gas. The LSR system velocity of the globule is about $-$23 km s$^{-1}$. The white arrow indicates the direction where the star cluster Tr 16 is located.  
    }
\label{HH1006_1}
\end{center}
\end{figure*}

\begin{table}[!]
\centering
\begin{tabular}{l*{7}{c}r}
\hline
\hline
\multicolumn{1}{c}{Source}&
\multicolumn{2}{c}{R \footnote{Radius of the pillar\textbackslash globule}}&
\multicolumn{2}{c}{$\dot{M}$\footnote{Mass photoevaporation rate}}&
\multicolumn{2}{c}{$t$\footnote{Photoevaphoration timescale}}\\
&{[au]} && [M$_{\odot}$ yr$^{-1}$]&   &[yr] & \\
\hline
\hline
HH 900&8050&&2.6$\times 10^{-5}$&&153&\\
HH 1006&8740&&5.7$\times10^{-6}$&&1228&\\
HH 1010&39100&&3.1$\times10^{-5}$&&2258&\\
\hline
\hline
\end{tabular}
\caption{The sizes of each pillar/globule were calculated using a Gaussian fitting in CASA to the extended gas emission using CO (2$-$1), assuming a distance of 2.3 kpc to each of the studied sources. Each pillar/globule is affected by ultraviolet radiation (F$_{EUV}$) originating from the nearby clusters, Tr 14 and Tr 16, for which a UV luminosity of approximately (Q$_H$) = 2$\times10^{50}$ photons s$^{-1}$ for Tr 14 and (Q$_H$) = 9$\times10^{50}$ photons s$^{-1}$ for Tr 16 \citep{s&b2008}.}
\label{table4}
\end{table}

\subsection{HH 1004}
This source is located in the southern region of the Carina Nebula, specifically within the South Pillars. The pillar identified by \citet{smith2010} exhibits a highly elongated structure pointing towards the Tr 16 stellar cluster. This displays a clear optical bipolar jet at the tip of the pillar, which is being strongly irradiated by the UV radiation field emanating from the OB stars present in Tr 16. Further south of the pillar, the presence of another bright jet is evident, expanding in a southwest direction opposite to that of the stellar cluster. There are shock structures further away that may potentially be associated with the same jet phenomenon, although considering only the innermost part of the detected optical jet, a jet length of approximately 0.4 pc is quantified. 

\begin{figure}[h]
\centering
\includegraphics[width=0.5\textwidth]{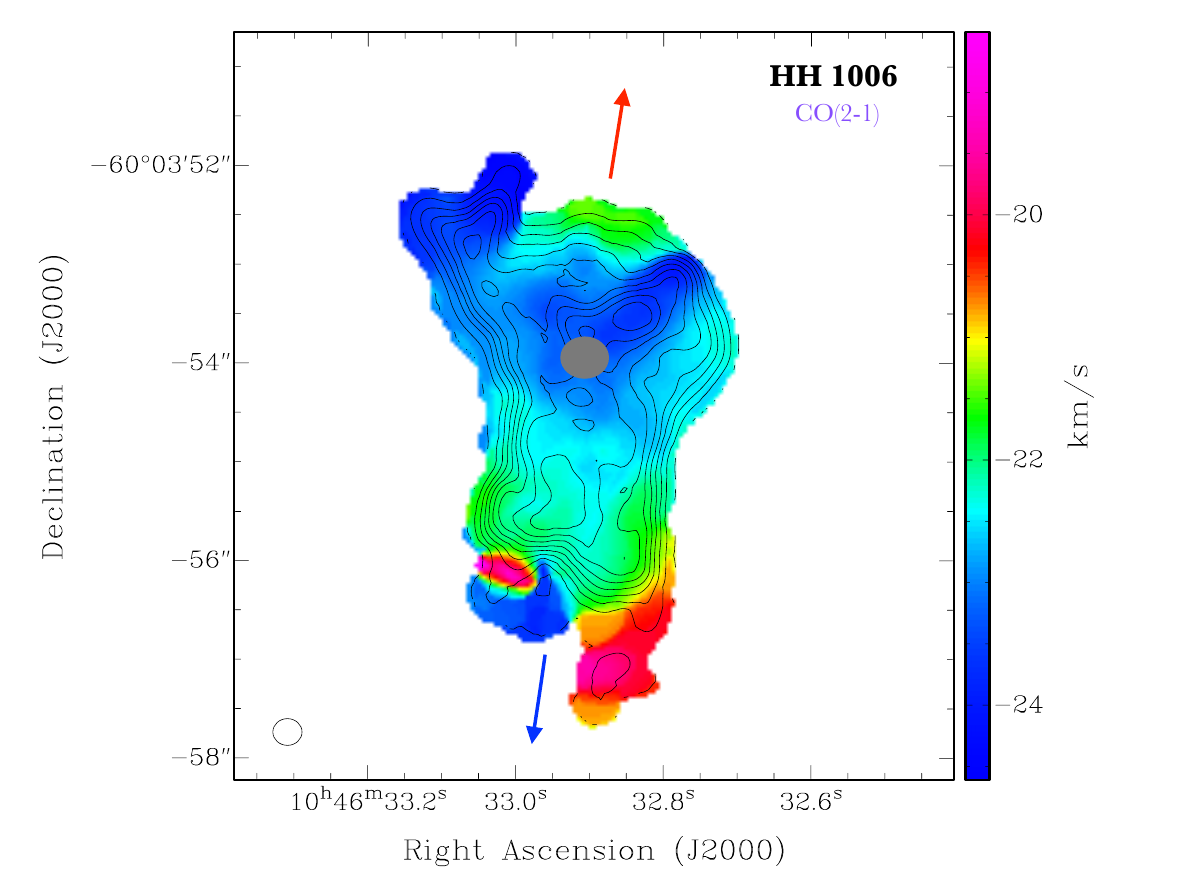}
\includegraphics[width=0.5\textwidth]{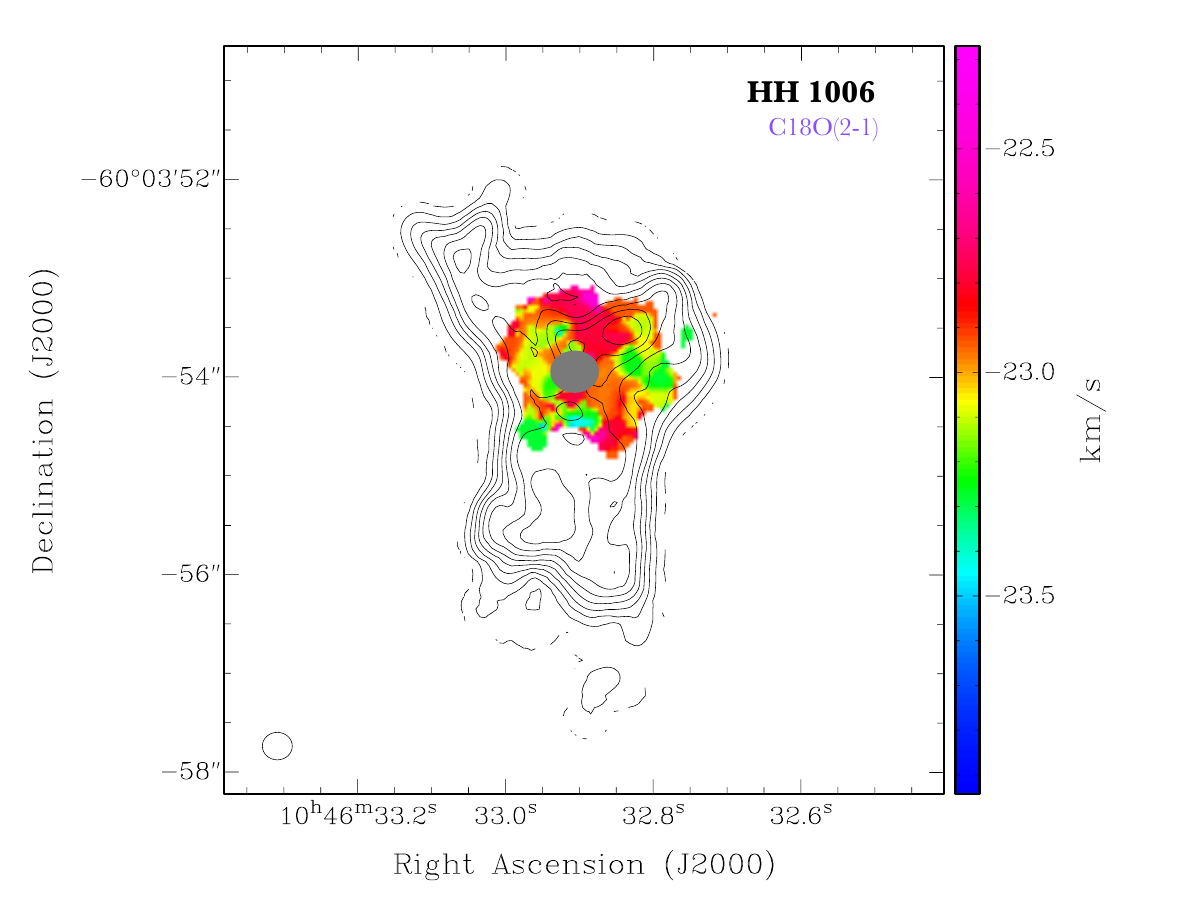}
\caption{Images of moment zero (contours) and moment one (colors) are presented for the molecular emission of CO (2$-$1) (top panel), and moment one of C$^{18}$O (2$-$1) (bottom panel) of the globule of HH 1006. The Figure presents a comparison between the CO moment zero presented in contours (varying in the range from 1\% to 90\% of the peak emission, in steps of 10\%), by the emission that is best tracing the globule, with respect to the moment one of emission for CO and C$^{18}$O. The peak of the millimeter line CO emission is 0.71 Jy Beam$^{-1}$ km s$^{-1}$. The arrows (blue and red) show the direction of molecular outflow expansion. The gray circles in both Figures represent the location of the compact source (system disk$+$envelope) detected in the continuum at 1.3 mm. The LSR radial-velocity scale bar is shown on the right.}  
\label{HH1006_2_3}
\end{figure}

In Figure \ref{hh1004_1}, the combination of observations made with {\it HST/ACS} for the HH 1004 pillar and our observations obtained with ALMA for the continuum and CO emission are shown. The optical image (green) corresponds to the H$\alpha$ emission and is overlapped with the ALMA high velocity CO(2$-$1) molecular emission (blue and red), and the 1.3 mm continuum emission (white contours). The CO(2$-$1) molecular emission was obtained by integrating in LSR velocities from $-$15.55 to $-$4.75 km s$^{-1}$ for the redshifted emission, and from $-$40.31 to $-$25.07 km s$^{-1}$ for the blueshifted emission. The ALMA 1.3 mm continuum image reveals three compact sources (A, B, and C) distributed in the northern region of the pillar (at the head of the pillar, see Table \ref{table1}). Components A and B appear to be associated with the optical jet detected in H$\alpha$. 

For this millimeter continuum emission, we obtained an integrated flux of 7.42 $\pm$ 0.38 mJy for source A, 0.90 $\pm$ 0.15 mJy for source B, and 1.49 $\pm$ 0.09 mJy for source C respectively. The observed rms noise was 0.053 mJy for the three sources, which is very close to the theoretical noise value (see Table \ref{table1}). From the synthesized beam size (0$^{\prime\prime}$.11 $\times$ 0$^{\prime\prime}$.06), we obtained a deconvolved size similar to HH 666, for example, for source C we obtained a deconvolved size of 0$^{\prime\prime}$.036 $\pm$ 0$^{\prime\prime}$.014 $\times$ 0$^{\prime\prime}$.017 $\pm$ 0$^{\prime\prime}$.018 with an PA of 178$^{\circ}$ $\pm$ 49$^{\circ}$, that corresponds to an approximate physical size of 83 au $\times$ 40 au, consistent with the typical physical sizes of the protoplanetary disks \citep{Andrews2020}. 

We estimated the total masses for these structures using the equation \ref{eq1} and we obtained that the masses for components A, B, and C are between 0.15$-$0.38, 0.01$-$0.04 and 0.03$-$0.07 M$_{\odot}$ respectively, assuming temperature values between 20 to 50 K (see Table \ref{table2}).

We show in the same Figure \ref{hh1004_1}, the molecular CO (2$-$1) emission plotted in blue and red colors corresponding to the blueshifted and redshifted gas, respectively.
We observe a bipolar outflow emerging from the compact continuum source A, with the blueshifted portion extending southwest along the axis of the optical jet detected in H$\alpha$. Very little redshifted emission is apparent, and due to the weak flux detected at these velocities, it is not clear whether this emission expands in the same northeast direction along the axis of the optical jet. Source B also appears to overlap a portion of the blueshifted molecular outflow detected, although it is difficult to discern whether the detection of this molecular outflow is generated by both sources A and B, or if it is solely attributed to source A, which is larger ($\sim$230 au $\times$ 140 au) and more massive (0.38 M$_{\odot}$) compared to source B ($\sim$140 au $\times$ 100 au, 0.04 M$_{\odot}$ mass). Compact source C exhibits a CO redshifted molecular outflow that is more extended than the one detected for sources A and B, although it does not display a blueshifted counterpart. This can be explained by the inhomogeneities in the cloud. As far as we know this is the first molecular detection for this outflow component. This source is north of the [Fe II] emission that goes through the pillar (see Fig 6 in \citet{reiter2016}). This looks to be in the right place to drive the extended H$\alpha$ emission seen off the left of the pillar that \citet{smith2010} identify as HH 1005.

\begin{figure*}[h!]
\begin{center}
\includegraphics[width=1.\textwidth]{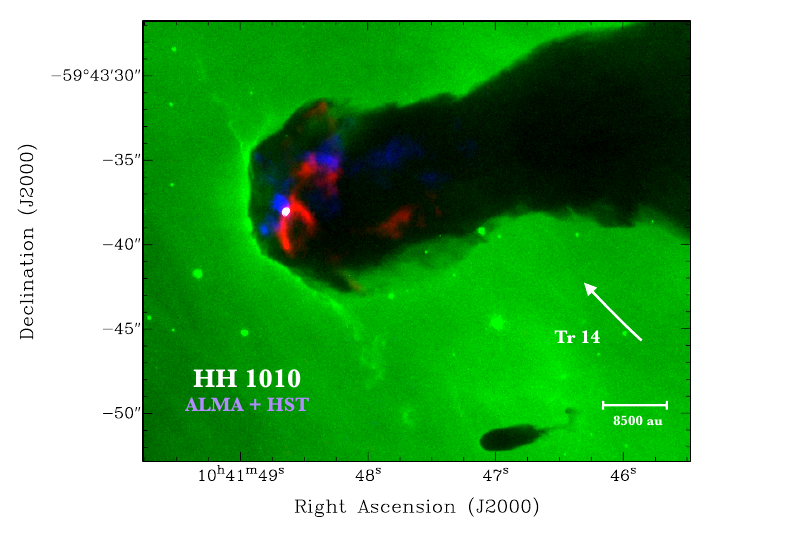}
\caption{ALMA CO(2$-$1) moment zero (blue and red colors) and millimeter continuum (white contours) images of the HH 1010 object overlaid in a HST (H$\alpha$) optical image (green colors). The blue and red colors represent the blueshifted and redshifted CO emissions, respectively, from the HH 1010 outflow. The 1.3 mm continuum emission is denoted by the white contours. The contour levels range from 18\% to 90\% of the peak emission, in steps of 10\%. The peak of the millimeter continuum emission is 4.07 mJy Beam$^{-1}$. Here, we are only contouring the most compact 1.3 mm emission from our observations revealing the envelope and the disk. We integrate in radial velocities from $-$9.54 to $-$15.25 km s$^{-1}$ for the redshifted emission, and from $-$18.43 to $-$27.32 km s$^{-1}$ for the blueshifted emission for the CO outflow in the HH 1010 object. The LSR system velocity of the entire pillar associated with the object HH 1010 is about $-$16.52 km s$^{-1}$.}
\label{hh1010_f1}
\end{center}
\end{figure*}

\begin{figure}[!h]
\centering
\includegraphics[width=0.5\textwidth]{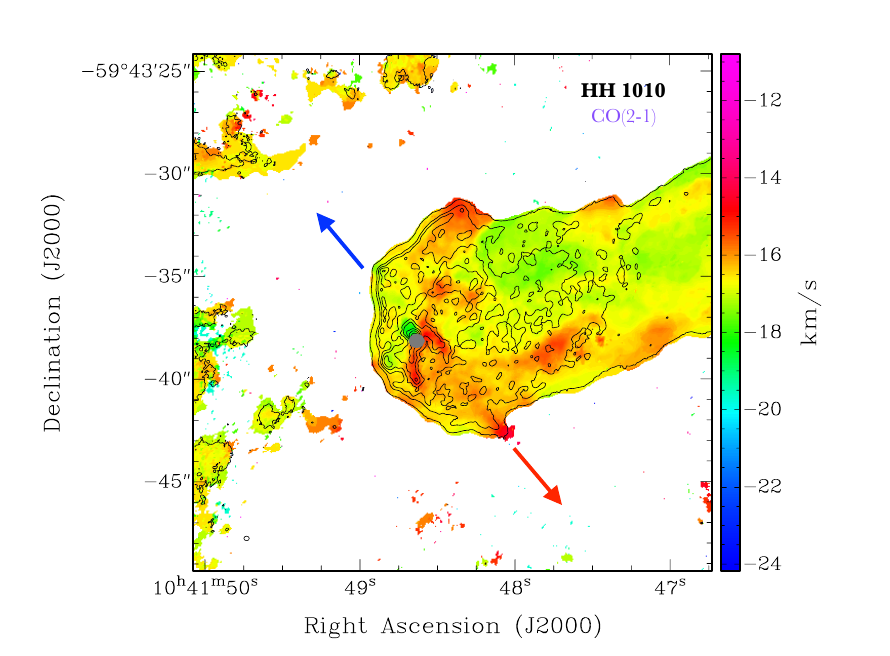}
\includegraphics[width=0.5\textwidth]{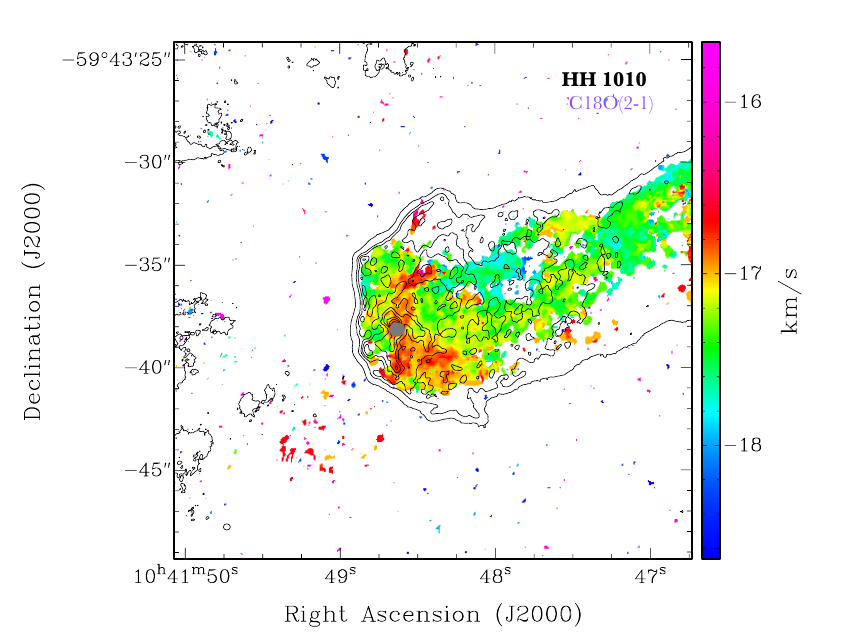}
\caption{Images of moment zero (contours) and moment one (colors) are presented for the molecular emission of CO (2$-$1) (top panel) and the moment one of C$^{18}$O (2$-$1) (bottom panel) of the globule of HH 1010. The Figure presents the CO moment zero emission in contours varying from 3\% to 90\% of the peak emission, in steps of 10\%. The peak of the millimeter line CO emission is 1.2 Jy Beam$^{-1}$ km s$^{-1}$. The arrows (blue and red) show the direction of molecular outflow expansion. The gray circles in both figures represent the location of the compact source (system disk$+$envelope) detected in the continuum at 1.3 mm. The LSR radial-velocity scale bar is shown on the right.}
\label{hh1010_f2}
\end{figure}

\subsection{HH 1006}
 This bipolar jet emerges from a small dark globule, pointing in the northern direction of $\eta$ Carinae, and is located in the central region of the South Pillars in the zone where these pillars are strongly irradiated by the massive stars of Tr 16. The HH 1006 jet is formed by a chain of knots that extend along its axis, and have an extension of $\sim$0.2 pc measured in optical wavelengths by \citet{smith2010}, and by \citet{sahai2012}, who presented CO (3$-$2) single dish observations, although for our study we have focused only on the characterization of the central jet, that is, the closest to the globule.

In Figure \ref{HH1006_1} we show an optical image of the HST in H$\alpha$ (green), and our results obtained with the ALMA observations for the continuum emission at 1.3 mm (white contours in the Figure) and the CO molecular emission (red and blue colors). The continuum emission reveals a compact source at the center of the globule with an integrated flux of 9.41 $\pm$ 0.73 mJy and a PA of 43$^{\circ}$ $\pm$ 17$^{\circ}$ (see Table\ref{table1}). From the synthesized beam size (0$^{\prime\prime}$.23 $\times$ 0$^{\prime\prime}$.20), we obtained a deconvolved size of 1$^{\prime \prime}$.139 $\pm$ 0.$^{\prime \prime}$089 $\times$ 0.$^{\prime \prime}$945 $\pm$ 0$^{\prime \prime}$.075, which corresponds to an approximate physical size of 2600 au $\times$ 2200 au. The compact source is generating the molecular outflow that extends along the same axis of the globule. We estimate that the mass of dust follows again the equation (\ref{eq1}) and is between 0.19$-$0.48 M$_\odot$, assuming temperature values between 20 K and 50 K. This value is an order of magnitude lower than the estimation of 0.02 M$_{\odot}$ obtained by \citet{mesa-delgado}, which can be explained by the fact that we report a larger source size. Perhaps, we are detecting more of the envelope than they did.
 
The molecular outflow is detected aligned on the axis of the optical jet and extending in the same direction of the globule axis at velocities between $-$21 and $-$24 km s$^{-1}$. This CO flux is tracing the innermost parts of the globule connected to the extent of the outflow with the driving source. We integrate in radial velocities from $-$17.88 to $-$21.05 km s$^{-1}$ for the redshifted emission oriented to the south, and from $-$24.23 to $-$27.40 km s$^{-1}$ for the blueshifted emission oriented to the north of the globule. 
 
In Figure \ref{HH1006_2_3}, we show the moment zero map obtained from CO(2$-$1) in contours overlaid with moment one map obtained from CO(2$-$1, upper panel) and C$^{18}$O(2$-$1, bottom panel), respectively.
It is observed that CO is perfectly tracing the globule hosting the compact source detected by the continuum (gray circle) and corresponding to the center of the bipolar molecular outflow also evident in the top panel. The width velocity is $\sim$ 5 km s$^{-1}$, which tells us that the molecular outflow is not high velocity one. The CO moment zero for the top panel image was obtained by integrating in the LSR velocity range of $-$17.88 to $-$27.40 km s$^{-1}$. The system LSR velocity of the cloud is $\sim$ $-$23 km s$^{-1}$. The CO is the best tracer of the globule, while the C$^{18}$O traces more internal zones of the cloud where the protostar and envelope are localized. The bottom panel image was obtained by integrating over the LSR velocity range of $-$21.93 to $-$24.60 km s$^{-1}$ with a width of velocities $\sim$ 1 km s$^{-1}$.

\begin{figure*}[!]
\begin{center}
    \includegraphics[width=0.99\textwidth]{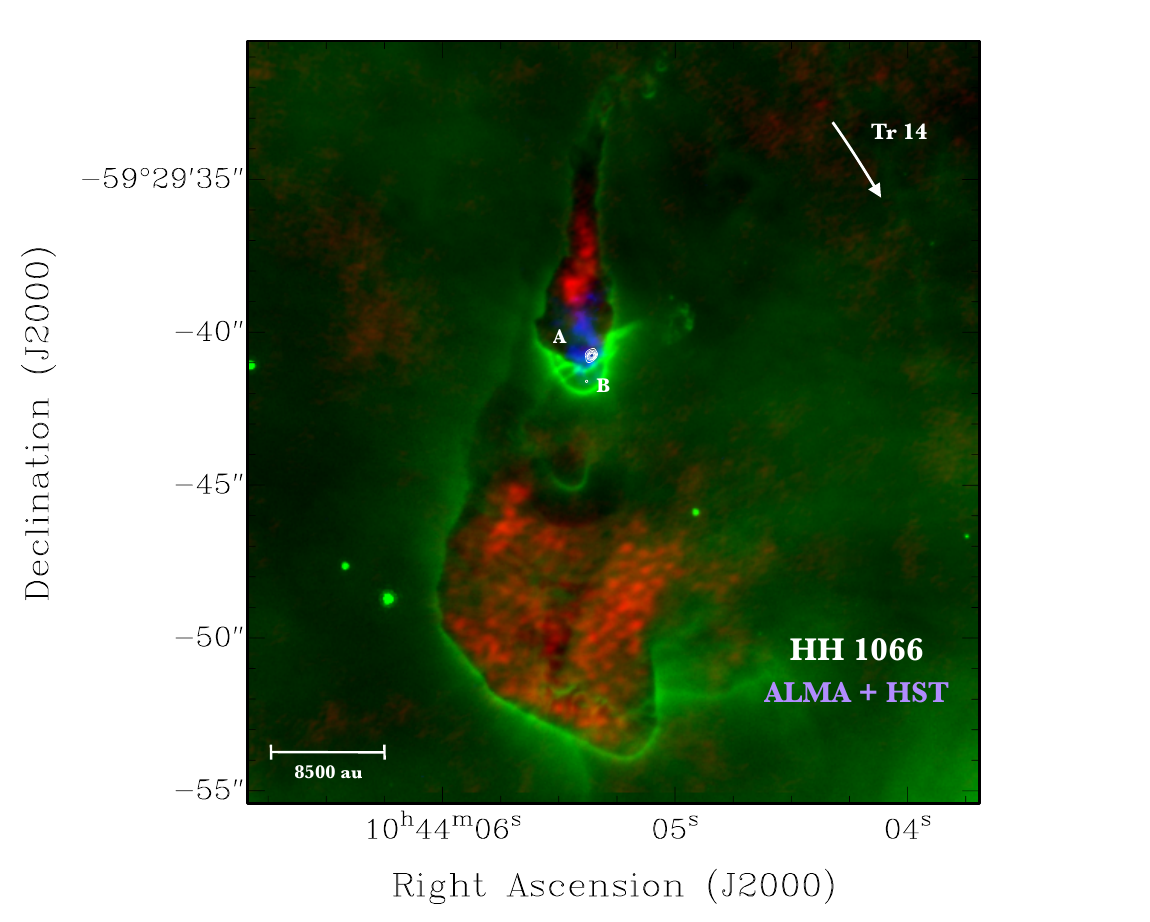}
    \caption{This image is a combination of a HST (optical) $+$ ALMA (radio) image for the globule HH 1066 located in {\it The Mystic Mountain} in Carina. The optical image (green) is the H$\alpha$ emission from the optical jet and obscured globule that houses the protostellar compact source traced by  ALMA 1.3 mm continuum emission (white contours varying from 50$\%$ to 90$\%$ of the peak emission, in steps of 10\%). The peak of the millimeter continuum emission is 1.96 mJy Beam$^{-1}$. We also show the C${18}$O (2$-$1) moment zero (blue and red colors) for the molecular outflow associated with the protostar. The blue color corresponds to the blueshifted molecular gas and the red color corresponds to the redshifted molecular gas. The LSR system velocity of the globule is about $-$9.18 km s$^{-1}$. The white arrow indicates the direction where the star cluster Tr 14 is located.
    }
\label{HH1066_1}
\end{center}
\end{figure*}

The gas mass of the globule is 4 $\times$ 10$^{-3}$ M$_{\odot}$ and the mass of the outflow is 2.1 $\times$ 10$^{-3}$ M$_{\odot}$ (see Table \ref{gasmass}). The globule HH 1006 is affected by the UV radiation coming from the cluster Tr 16, which has a UV luminosity of (Q$_H$) = 9$\times10^{50}$ photon s$^{-1}$, and given the distance at which the source of the cluster is located ($\sim$16 pc), we obtain a flux F$_{EUV}=3.09\times10^{10}$ photons s$^{-1}$ cm$^{-2}$. We obtained a mass photoevaporation rate of 5.7$\times 10^{-6}$ M$_{\odot}$ yr$^{-1}$ and the photoevaporation timescale of the globule HH 1006 given the estimated mass of gas (7 $\times$ 10$^{-3}$ M$_{\odot}$) is $\sim$1230 yr (see Table \ref{table4}).

\subsection{HH 1010}
This source emerges from the tip of a giant dark pillar located at the western periphery of the Carina nebula, and pointing toward the evolved star $\eta$ Carinae and Tr 16. \citet{smith2010} detected optical emission within the pillar, although their detection could not be corroborated. From the optical images, the presence of a bipolar jet becomes evident, also reported by \citet{mcleod2016}, with its jet axis perpendicular to the pillar axis. This geometrical configuration is commonly observed in Carina Pillars.

In Figure \ref{hh1010_f1}, we show an optical image of the HST in H$\alpha$ (green), and our results obtained with the ALMA observations for the continuum emission at 1.3 mm (white contours) and the CO (2$-$1) molecular emission (red and blue colors). The continuum emission reveals a compact source at the center of the pillar head, and that is exciting the molecular outflow. This outflow extends along the jet axis at velocities between $-$9 and $-$27 km s$^{-1}$. This continuum emission has an integrated flux of 4.12 $\pm$ 0.17 mJy. We were not able to resolve the source because the angular resolution was not sufficient (see Table \ref{table1}). 

The dust mass for the continuum source is between 0.08$-$0.21 M$_\odot$, assuming temperature values between 20 and 50 K. The CO is tracing the outflow and was obtained by integrating the emission in radial velocities from $-$9.54 to $-$15.25 km s$^{-1}$ for the redshifted emission oriented to the south, and from $-$18.43 to $-$27.32 km s$^{-1}$ for the blueshifted emission oriented to the north of the pillar. Red/Blueshifted distribution is consistent with \citet{mcleod2016} observations. 

\begin{figure}[!h]
\centering
\includegraphics[width=0.5\textwidth]{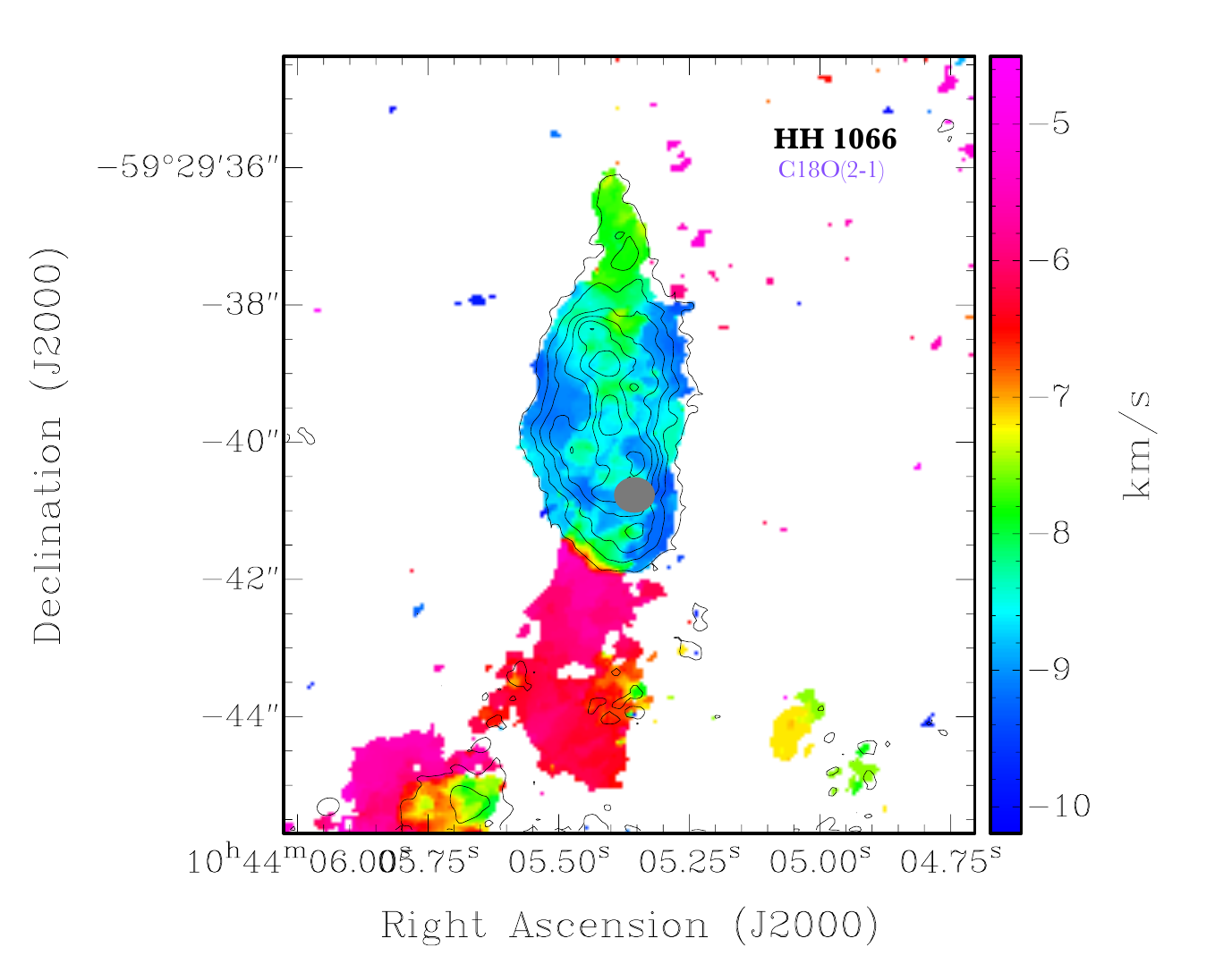}
\includegraphics[width=0.5\textwidth]{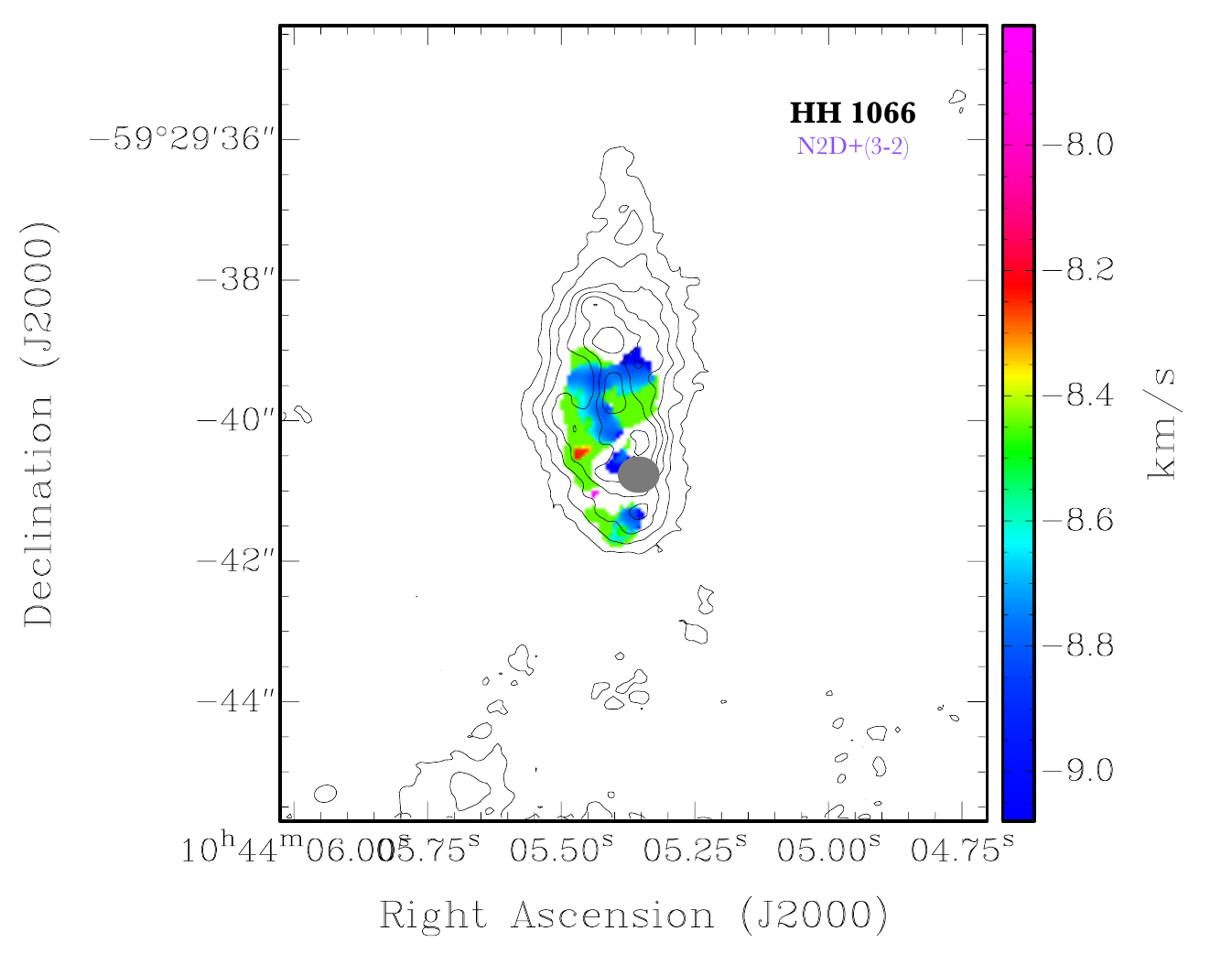}
\caption{Images of moment zero (contours) and moment one (colors) are presented for the molecular emission of C$^{18}$O (2$-$1) (top panel) and moment one of N$_2$D$^{+}$(3$-$2) (bottom panel) of the globule of HH 1066. The C$^{18}$O moment zero presents the emission in contours varying from 10\% to 90\% of the peak emission, in steps of 15\%. The peak of the millimeter line C$^{18}$O emission is 0.11 Jy Beam$^{-1}$ km s$^{-1}$. The gray circles in both figures represent the location of the compact source (system disk$+$envelope) detected in the continuum at 1.3 mm. The LSR radial-velocity scale bar is shown on the right.}
\label{hh1066_f2}
\end{figure}

In Figure \ref{hh1010_f2}, we show the emission of molecular gas for the CO (2$-$1) and C$^{18}$O (2$-$1) transitions. The moment zero (black contours) represents the CO emission in both panels. The moment one is presented (in colors) for the CO (top panel) and C$^{18}$O (bottom panel). For the top panel, the CO moment zero image was obtained by integrating in the velocity range of $-$9.54 to $-$27.32 km s$^{-1}$, with the average LSR velocity of the cloud $\sim$ $-$16 km s$^{-1}$. The bottom panel image was obtained by integrating over the velocity range of $-$15.65 to $-$18.65 km s$^{-1}$ km s$^{-1}$, with the average LSR velocity of the cloud $\sim$ $-$17 km s$^{-1}$. We note the fact of the CO is tracing perfectly the pillar in all its extension and the accurate shape of the molecular outflow is still evident even considering all the CO emission. The C$^{18}$O is again tracing zones deep inside the cloud, where the protostar is forming and therefore the density is higher. 

The estimated gas mass for the HH 1010 object is 7 $\times$ 10$^{-2}$ M$_{\odot}$ and the mass of the outflow is 1.5 $\times$ 10$^{-3}$ M$_{\odot}$ (see Table \ref{gasmass}). The globule HH 1010 is affected by the UV radiation coming from the cluster Tr 14, given the distance at which the source of the cluster is located ($\sim$13 pc), we obtain a flux of F$_{EUV}=1.02\times10^{10}$ photons s$^{-1}$ cm$^{-2}$, and the mass photoevaporation rate 3.1$\times 10^{-5}$ M$_{\odot}$ yr$^{-1}$, yielding a photoevaporation timescale $\sim$2260 yr (see Table \ref{table4}).

\subsection{HH 1066}

This object was detected to the west of the Carina nebula very close to the Tr 14 star cluster just above the HH 901 and 902 sources, which all together form the structure known as {\it The Mystic Mountain\footnote{https://hubblesite.org/contents/media/images/2010/13/2707-Image.html?news=true}}. The detected jet is approximately 0.05 pc in extent and is associated with what appears to be a small, bright bow shock (see \citeauthor{smith2010} \citeyear{smith2010}) although without an obvious optical jet counterpart. Through optical observations (H$\alpha$) they were able to detect a weak point source of emission located at the tip of the dark cloud and within the axis of the optical jet, which was also detected with IR observations of \textit{Spitzer} (YSO detected along the optical jet, see \citeauthor{povich2011} \citeyear{povich2011}) without a clear association with the driving source. Further optical and IR observations \citet{Reiter2013} made evident the bipolarity of the collimated jet detected in H$\alpha$ and [Fe II] but again without being able to evidence the complementary bow shock.

In Figure \ref{HH1066_1}, we show an optical image from the HST (in H$\alpha$, in green) and our results obtained with ALMA observations for the continuum emission (at 1.3 mm in white contours) and the C$^{18}$O molecular emission (red and blue colors). In this case, it was not possible to highlight the outflow with CO(2$-$1) emission due to contamination of ambient gas, and the structures near the globule. Although we tried to reproduce the innermost zone (blue and red colors) of the C$^{18}$O molecular outflow (in the previous cases where we have used CO), the emission plotted in Figure \ref{HH1066_1} only represents the innermost zones of the molecular cloud, without clearly outlining the outflow. The emission in this case was integrated over the LSR velocity range of $-$7.51 to $-$8.18 km s$^{-1}$ for the redshifted gas and $-$9.84 to $-10.18$ for the blueshifted gas.

From the continuum emission, we detected a compact dust source associated with the envelope and disk structure, where the protostellar source is embedded (component A, see Figure 9), and a compact emission source (component B). For component A, we obtained an integrated flux of 3.86 $\pm$ 0.66 mJy, and a rms noise of 0.05 mJy (see Table \ref{table1}). From the synthesized beam size (0$^{\prime\prime}$.26 $\times$ 0$^{\prime\prime}$.18), we obtained a deconvolved size of 0.$^{\prime \prime}$373 $\pm$ 0.$^{\prime \prime}$085 $\times$ 0.$^{\prime \prime}$211 $\pm$ 0.$^{\prime \prime}$092, with a PA of 170$^{\circ} \pm$ 26$^{\circ}$, which corresponds to an approximate physical size of 850 $\times$ 490 au. We estimated the dust mass from the continuum emission and obtained values between 0.07$-$0.19 M$_{\odot}$, varying the temperature between 20 and 50 K, these values are in agreement with the estimated values by  \citet{mesa-delgado} (they estimated the mass $\sim$ 0.04M$_{\odot}$), we clearly are recovering some extended emission. Additionally, we report for the first time the detection for component B, whose integrated flux of 2.9 mJy and peak flux of 0.8 mJy, are about an order of magnitude larger than rms value. There was no gas detection associated with this component.  

In Figure \ref{hh1066_f2}, we show the moment zero obtained from the C$^{18}$O (2$-$1) (contours) combined with moment one image obtained from the C$^{18}$O (2$-$1) (upper panel) and N$_2$D$^{+}$(3$-$2) (bottom panel) emissions, respectively. This is the only source in this study where we detected N$_2$D$^{+}$(3$-$2) emission. The moment zero image of C$^{18}$O perfectly traces the globule of HH 1066, and it does not extend beyond the boundary delimited by the optical emission detected in H$\alpha$ (see Figure \ref{HH1066_1}). The moment one reveals gas kinematics, that for C$^{18}$O, evidences a small velocity gradient with ranges between $-$7 to $-$10 km s$^{-1}$, with higher velocities toward the outer walls of the globule. In the case of N$_2$D$^{+}$ moment one, the velocities vary over a smaller range from $-$8 to $-$9 km s$^{-1}$, and the structure traced by the gas is located in the innermost regions of the globule, close to the protostellar source detected by the continuum emission (gray circle). 

 \subsection{Planet Formation}
We can examine if the results obtained for the compact disks$+$envelopes revealed in our 1.3 mm continuum maps for the HH objects are in principle consistent with the formation of planets. If we consider the lowest value obtained for the masses of the disks$+$envelopes (0.01 M$_{\odot}$ for HH 1004 component B) which corresponds to $\sim$ 10 M$_{Jup}$, we can see that is above the minimum required mass (10 M$_{Jup}$) for a pre-solar nebula to develop a planetary system \citep{Wei1977}. However, we remark that there is probably contamination of the dusty envelope, so the mass of the disk could be lower. However, if we consider the maximum mass value obtained (0.77 M$_{\odot}$ for HH 666), corresponding to $\sim$ 800 M$_{Jup}$, we see that it is well above the minimum required, which is the case for many other sources detected with our observations, so at least in most of the sources this minimum mass condition is fulfilled.

On the other hand, a second important factor to consider is the average age of the Carina population which is $\sim$1–4 Myr \citep{s&b2008}, consistent with the minimum time-scale required to form planets $\sim$1–2 Myr (see \citet{hubickyj2005,Najita2014}). The external UV irradiance from Tr 14 and Tr 16 could photodisociate the material of the disks. However, these disks are still very embedded within the globules and pillars implying that they are not strongly irradiated yet. All these conditions indicate the possibility that there are young planets formed or in formation in these detected disks. 

\section{Conclusions}
In this work, we present a study of six pillars/globules localized in the star forming region of Carina. We used Atacama Large Millimeter/submillimeter Array observations at 1.3 mm continuum, C$^{18}$O(2$-$1) and $^{12}$CO(2$-$1) spectral lines with high angular resolution and sensitivity. We have studied each source associated with the Herbig-Haro objects, and their nearby environment, aiming to identify and analyze various characteristics that are influencing each of the pillars/globules. The most significant results derived from this study are as follows:

\begin{enumerate}\renewcommand{\theenumi}{\roman{enumi}}
    \item For the first time, we have successfully identified the  millimeter compact sources associated with the Herbig-Haro objects HH 666, HH 1004, HH 1006, HH 1010, and HH 1066 along with the first detection of its component B. We also detected the CO bipolar molecular outflows associated with these continuum sources, extending along the optical jet axis.
    \item From the observations, we have characterized the objects from the morphological analysis of the detected structures, revealing sizes of the pillars/globules and of the detected continuum structures, for which we have concluded that, given the sizes found (80$-$2600 au), they correspond to structures associated with the dusty disks ($\sim$100 au) and the envelope of the forming star, for which we have also calculated the dust masses, finding values between 0.01$-$0.7 M$_{\odot}$. Between these values, planetary formation would be possible.
    \item We have also estimated the gas mass for the detected molecular outflows from the CO emission, giving values of 1.4$\times$10$^{-3}$ $-$ 2.1$\times$ 10$^{-3}$ M$_{\odot}$, and the same for the pillars/globules, which obtained values of 7$\times$10$^{-3}$ $-$ 7$\times$10$^{-2}$ M$_{\odot}$. These estimations only include the HH 900, HH 1006, HH 1010, and HH 1066 sources since only for these we obtained CO detection for the pillars/globules.
    \item We present the gas kinematics for the $^{12}$CO and C$^{18}$O molecular lines, estimations for the mass photoevaporation rates for the HH 900, HH 1006, HH 1010, and HH 1066 sources, with a range from 5.7 $\times$ 10$^{-6}$ to 3.1 $\times$ 10$^{-5}$ M$_{\odot}$ yr$^{-1}$, indicating that these pillars/globules will be photoevaporated on time scales of the order of 150$-$2200 yr.
\end{enumerate}

\begin{acknowledgments}
 This paper makes use of the following ALMA data: ADS/JAO.ALMA\#2017.1.00912.S ALMA is a partnership of ESO (representing its member states), NSF (USA), and NINS (Japan), together with NRC (Canada), MOST and ASIAA (Taiwan), and KASI (Republic of Korea), in cooperation with the Republic of Chile. The Joint ALMA Observatory is operated by ESO, AUI/NRAO, and NAOJ.  The National Radio Astronomy Observatory is a facility of the National Science Foundation operated under cooperative agreement by Associated Universities, Inc.
G.C.R. is grateful to CONACyT, México, and DGAPA, UNAM for the financial support.
L.A.Z. acknowledges financial support from CONACyT-280775 and UNAM-PAPIIT IN110618, and IN112323 grants, México. P.R.R-O. acknowledges support from the UNAM-PAPIIT IN110722 grant, CONACyT, México, and DGAPA, UNAM.
The HST data presented in this paper were obtained from the Mikulski Archive for Space Telescopes (MAST) at the Space Telescope Science Institute. The specific observations analyzed can be accessed via \dataset[10.17909/p1je-9b56]{https://doi.org/10.17909/p1je-9b56}.
\end{acknowledgments}

\vspace{5mm}
\newpage
\bibliography{sample631}{}
\bibliographystyle{aasjournal}

\end{document}